\documentclass[a4paper,11pt]{article}
\usepackage{jcappub} 
\usepackage{aas_macros} 
\usepackage{lineno}

\usepackage{comment}
\usepackage{appendix}
\title{\boldmath BBN bounds on neutrinophilic ultralight Dark Matter}

\author[a]{T. Bertólez-Martínez}
\author[b]{J. López-Sarrión}
\author[a]{J. Salvado}
\affiliation[a]{Departament de Física Quàntica i Astrofísica and Institut de Ciències del Cosmos, Universitat de Barcelona, Diagonal 647, E-08028 Barcelona, Spain}
\affiliation[b]{Departamento de Física Teórica and Centro de Astropartículas y Física de Altas Energías (CAPA), Universidad de Zaragoza, Zaragoza 50009, Spain}

\emailAdd{antoni.bertolez@fqa.ub.edu}
\emailAdd{justo.lopezsarrion@gmail.com}
\emailAdd{jsalvado@icc.ub.edu}

\newcommand{\dpppE}{\frac{d^3\vec{p}}{(2\pi)^3}\frac{1}{\sqrt{2E_p}}}
\newcommand{\dd}{\mathrm{d}}

\abstract{
The high densities in the early Universe provide a unique laboratory to constrain couplings between feebly interacting particles, such as dark matter and neutrinos. In this article, we study how Big Bang Nucleosynthesis can constrain models of Ultra-Light Dark Matter diagonally coupled to neutrinos. We follow an adiabatic formalism which allows to average-out the rapid oscillations of the Dark Matter field and consistently take into account the feedback between the neutrino and the Dark Matter fields. This feedback alters the early Universe dynamics, causing the Dark Matter energy density to scale as radiation, while the neutrino mass scales as $a^{-1}$. 
These two effects modify primordial element abundances by modifying interaction rates and the expansion rate during nucleosynthesis. 
Then, we use primordial abundances to obtain leading cosmological bounds on the coupling in the range $m_\phi/{\rm eV}\in (10^{-22},10^{-17})$, namely $g\lesssim 0.13(m_\phi/{\rm eV})$ for $m_\phi \gtrsim 3\times 10^{-20}\,\rm eV$ and $g\lesssim 1.8\times 10^{-11}\sqrt{m_\phi/\rm eV}$ for $m_\phi \lesssim 3\times 10^{-20}\,\rm eV$. This consistent cosmological treatment emphasizes that, in the mass interval where its physical assumptions hold, neutrino masses cannot be generated refractively by a direct coupling with an Ultra-Light Dark Matter field.
}

\begin{document}
\maketitle
\flushbottom

\section{Introduction}\label{sec:intro}
The nature and properties of Dark Matter (DM) remain elusive.
In the hope of shedding light into the darkness, particle physics, astrophysics, and cosmology have managed to greatly constrain the portals that can connect the dark sector of DM with the SM~\cite{Cirelli:2024ssz,OHare:2024nmr,Arcadi:2024ukq,Fabbrichesi:2020wbt}. One of such possibilities is through the neutrino portal, i.e., the possibility that DM is neutrinophilic and only interacts with the SM through neutrinos. For instance, on the theory side such interactions are motivated by many models which predict a connection between the origin of non-zero neutrino masses and the nature of DM~\cite{Tao:1996vb,Ma:2006km}. However, due to the elusive nature of neutrinos, current experimental and observational constraints on $\nu$-DM couplings are much weaker than their electromagnetic counterparts~\cite{AxionLimits}. This motivates the search for new phenomenological approaches which can constrain them. 

One particularly interesting candidate to DM are light scalar fields, which are predicted by many extensions to the SM~\cite{Svrcek:2006yi,Green:1987mn,Arvanitaki:2009fg,Dine:2007zp,Halverson:2017deq,Bachlechner:2018gew,Essig:2013lka}. Since they could be non-thermally produced through the misalignment mechanism --where the initial (random) value of the field is displaced from zero--~\cite{OHare:2024nmr,Preskill:1982cy,Abbott:1982af,Dine:1982ah,Turner:1983he}, these bosons could make up for the entirety of DM. When these bosons have mass $m_\phi \ll 1\, \mathrm{eV}$, namely $m_\phi\sim 10^{-22}-10^{-10}\, \mathrm{eV}$, they are called Ultra-Light Dark Matter (ULDM) candidates, and have their own distinct phenomenology~\cite{Hui:2016ltb,Hui:2021tkt, Marsh:2015xka,Hu:2000ke,OHare:2024nmr,Sin:1992bg,Peebles:2000yy,Press:1989id,Turner:1983he,Baldeschi:1983mq,Goodman:2000tg,Arbey:2003sj,Amendola:2005ad,Chavanis:2011uldm,Suarez:2011yf,Berezhiani:2015bqa,Fan:2016rda}. 

The unconstrained properties of neutrinos and DM make $\nu$-ULDM interactions span a wide range of interesting phenomenology, from distorted neutrino oscillations to high-energy astrophysics~\cite{Blennow:2019fhy,Berlin:2016woy,Dev:2020kgz,Krnjaic:2017zlz,Barger:2005mn,Brdar:2017kbt,Losada:2021bxx,Capozzi:2018bps,Cheek:2025kks,ChoeJo:2023ffp,Chun:2021ief,Choi:2019zxy,Liao:2018byh,Ge:2019tdi,Murgui:2023kig,Ge:2024ftz,Reynoso:2016hjr,Reynoso:2022vrn,Lambiase:2023hpq,Pandey:2018wvh,Farzan:2018pnk,Huang:2018cwo,Huang:2021kam,Huang:2022wmz,Dev:2022bae,Plestid:2024kyy,Venzor:2020ova,Davoudiasl:2018hjw,Smirnov:2021zgn,Martinez-Mirave:2024dmw,Kim:2025xum,Babu:2019iml,Sen:2023uga,Sen:2024pgb,Lee:2024rdc,Ma:2006fn,Berlin:2016bdv,Lambiase:2025twn}. 
One of the main interests of these interactions is that neutrinos acquire a mass by propagating through a medium of ULDM. Hence, these models might be able to provide a joint explanation for neutrino masses and dark matter~\cite{Sawyer:1998ac,Berlin:2016woy,Choi:2020ydp,Sen:2023uga}. Since the DM number density increases in the early Universe as $a^{-3}$, these models predict a varying neutrino mass with $m_\nu \sim a^{-3/2}$. 
If one then requires neutrino masses to have the right order of magnitude in terrestrial oscillation experiments, then this predicts $m_\nu \sim 10\, \mathrm{eV}$ in the CMB epoch. As discussed in~\cite{Sen:2023uga,Sen:2024pgb}, this naive estimate motivates that the interaction is mediated by a light fermion mediator, which allows surpassing cosmological bounds. 

However, since cosmology is not directly sensitive to the neutrino mass~\cite{Bertolez-Martinez:2024wez}, this naive estimate must be refined in order to better understand when are such mediators necessary.
To achieve that, one must understand the coupled evolution of neutrinos and ULDM, and follow the evolution of its energy density and pressure. For instance, if neutrinos acquire a mass when the amplitude of ULDM increases, this is an energy expense which modifies the dynamics of ULDM~\cite{Plestid:2024kyy,Huang:2021kam}. As shown in~\cite{Huang:2021kam}, this modifies the neutrino mass scaling to $m_\nu\sim a^{-1}$.

In this work, we consistently study the cosmological implications of a coupling between Dirac neutrinos ($\psi$) and the ULDM field ($\phi$). We do so through a pseudoscalar coupling, $ig\bar\psi \gamma^5\phi\psi$~\cite{Moody:1984ba,Kim:1986ax}, which predicts mass-varying neutrinos with no coherent long-range interactions; but most results also apply to an scalar coupling. Then, we quantitatively compute the background evolution of the coupled field, accounting for the fast oscillations of the ULDM with an adiabatic approximation. Apart from modifying the scaling of $m_\nu$, this full calculation predicts additional relativistic degrees of freedom in the early Universe. The effect of these two phenomena on primordial element abundances from Big Bang Nucleosynthesis can constrain the properties of the coupled $\nu$-ULDM fluid and, thus, their coupling.

The outline of this article is as follows. In \cref{sec3:pseudo-formalism}, we introduce the action for the model and derive the equations of motion for both species. In \cref{sec3:pseudo-adiabatic}, we introduce the adiabatic approximation, a necessary ingredient to solve the equation of motion for an ULDM field. In \cref{sec3:pseudo-phenomenology}, we describe the cosmological implications of this model and its phenomenological signatures. In particular, in \cref{sec3:pseudo-bbn} we discuss its effects on Big Bang Nucleosynthesis (BBN). Finally, in \cref{sec3:pseudo-results} we present our results and in \cref{sec3:pseudo-conclusions} we conclude.

\section{Formalism}\label{sec3:pseudo-formalism}
We consider the following action in a curved spacetime background~\cite{Parker_Toms_2009}
\begin{equation}\label{eq3:lagrangian-pseudo}
\begin{split}
  S &= \int \sqrt{-\det g}\,  \dd^4 x\, \mathcal{L}(x) = \\ &=
  \int \sqrt{-\det g}\,  \dd^4 x \left(
  -\frac{1}{2}D_\mu \hat\phi\, D^\mu \hat\phi - \frac{1}{2}m_\phi^2\hat\phi^2 
                + \bar{\psi}\left[i\slashed{D}-m_0 + ig\hat\phi\gamma^5\right]\psi\right) \, .
\end{split}
\end{equation}
Here, $\mathcal{L}$ is the Lagrangian density of the model, $D_\mu$ is a covariant derivative, $\det g$ is the metric determinant, $\bar\psi = \psi^\dagger\gamma^0$, $m_\phi$ is the bare mass of the pseudoscalar field, $m_0$ the bare mass of the neutrino field and $g$ the coupling between both fields. Naively, gauge invariance asks that a coupling must exist both to electrons and to neutrinos. However, the coupling to electrons can be suppressed, for instance, if the ultralight pseudoscalar is coupled to an SM-singlet right-handed neutrino~\cite{Berlin:2016woy,Krnjaic:2017zlz}. Then, in this work $\psi$ can safely refer to active neutrinos only. As a first step, we choose a diagonal coupling, which will not lead to modified neutrino oscillations. 

From~\cref{eq3:lagrangian-pseudo}, the equations of motion for the quantum fields are
\begin{align}
  \label{eq3:eom-pseudoscalar}
  -D^\mu D_\mu\hat\phi + m_\phi^2\hat\phi =\ & ig \bar\psi\gamma^5\psi\, , \\
  \label{eq3:eom-fermion-pseudo}
  i\slashed{D}\psi - (m_0-ig\hat\phi\gamma^5)\psi =\  & 0\, .
\end{align}
Within a cosmological framework, the occupation number of the pseudoscalar field $\hat\phi$ is large, and it can be described by a coherent state. 
The fermion field $\psi$ can be described as an ensemble of classical particles with a given momentum $\vec{p}$. In this framework, every quantum operator $\hat{\mathcal{O}}$ is replaced by its expectation value~\cite{Esteban:2021ozz}
\begin{equation}\label{eq3:expectation-value}
  \langle\hat{\mathcal{O}}\rangle = 
  \sum_s \int \dd P_1\dd P_2 \dd P_3 \frac{1}{\sqrt{-\det g}}
  \frac{1}{2P^0} f(P,x,s) \langle\phi,P^s| \hat{\mathcal{O}}|\phi,P^s\rangle\, .
\end{equation}
$P^\mu$ are the conjugate momenta to the positions $x^i$ which solve the geodesic equation, $s$ is the spin and $f(P,x,s)$ the distribution function. $|\phi,P^s\rangle = |\phi\rangle \otimes|P^s\rangle$ is the product of a coherent state with field $\phi$ and a one-particle fermion state. \Cref{app:formalism} gathers the formal definitions for these states. 

\subsection{Fermionic equation of motion}
We first analyze the equation of motion of fermions, \cref{eq3:eom-fermion-pseudo}, where for now we treat $\phi$ as a constant and homogeneous value. As discussed later, this is a very good approximation for ultralight fields as the one under consideration. We perform the chiral rotation
\begin{equation}
  \psi \to e^{i\alpha\gamma^5}\psi = 
       \left(\cos\alpha + i\gamma^5\sin\alpha\right)\psi\, ,
\end{equation}
where $\alpha$ is a constant parameter. The equation now becomes
\begin{equation}
  \left(i\slashed{\partial}-(m_0\cos\alpha - g\phi\sin\alpha)
          +i\gamma^5(-m_0\sin\alpha+g\phi\cos\alpha)\right)\psi = 0\, ,
\end{equation}
where we have used $(\gamma^5)^2 = \mathbb{I}_4$, the 4x4 identity matrix. Now, by making the right choice,
\begin{equation}
  \tan 2\alpha = \frac{g\phi}{m_0}\, ,
\end{equation}
we convert this equation into
\begin{equation}\label{eq3:eom-fermion-diagonalised}
  \left(i\slashed{\partial}-m_\nu\right)\psi = 0\, .
\end{equation}
This is the standard Dirac equation with a mass term given by
\begin{equation}\label{eq3:fermion-effective-mass}
  m_\nu = \sqrt{m_0^2 + g^2\phi^2}\, .
\end{equation}
Now, this tells us that the eigenstates to the Dirac operator from \cref{eq3:eom-fermion-pseudo} are two states with the same mass, $m_\nu$. The same result would have been achieved by directly doing the change of variables in the Lagrangian.
The plane wave solutions to~\cref{eq3:eom-fermion-pseudo} are
\begin{equation}
  u^s(\vec{p}) = e^{i\alpha \gamma^5}u^s_0(\vec{p})\,,\qquad
  v^s(\vec{p}) = e^{i\alpha \gamma^5}v^s_0(\vec{p})\, ,
\end{equation}
where $u^s_0(\vec{p}), v^s_0(\vec{p})$ are the plane wave solutions for a free fermion of mass $m_\nu$. These solutions fulfill,
\begin{equation}
    \bar{u}^s(\vec{p})\gamma^5 u^r(\vec{p}) = 
    -\bar{v}^s(\vec{p})\gamma^5 v^r(\vec{p}) = 2ig\phi\, \delta^{sr}\, .
\end{equation}
The chiral rotation of the Dirac eigenstates rotates the electroweak vertex with a global phase and thus scattering amplitudes are not modified.

Two main differences appear if neutrinos are coupled to the ULDM field through a scalar term $g\bar\psi\phi\psi$. On the one hand, the effective neutrino mass has a different dependence, $m_\nu = m_0+g\phi$~\cite{Esteban:2021ozz,Smirnov:2022sfo,Gao:2021fyk}. However, in the early Universe, neutrino mass is dominated by $g\phi$, and so the scalar and pseudoscalar equally predict $m_\nu^2 \sim g^2\phi^2$. For example, for $m_0\sim 10^{-3}\, \rm eV$, this happens at redshift $z\gg 150$.
On the other hand, long-range self-interactions in the pseudoscalar coupling are spin-dependent and thus irrelevant, while self-interactions in the scalar coupling are important when $gm_0/m_\phi \gtrsim 10$~\cite{Esteban:2021ozz}. For $m_0 \lesssim 10^{-3}\, \rm eV$, couplings probed here make self-interactions also negligible in the scalar formalism (see \cref{fig3:pseudo-money-plot}).
As a consequence, quantitative results shown in \cref{sec3:pseudo-results} also apply to the scalar coupling. However, the formalism introduced here is not applicable to the scalar coupling at late time, where the mass dependence is different.

\subsection{Pseudoscalar equation of motion}
Now, we take the expectation value of~\cref{eq3:eom-pseudoscalar} assuming an homogeneous background field $\phi = \phi(t)$ and a collection of particles as in \cref{eq3:expectation-value}, which leads to 
\begin{equation}
  -D_\mu D^\mu\phi + m_\phi^2\phi = ig\sum_s\int \dd P_1\dd P_2\dd P_3
  \frac{1}{\sqrt{-\det g}}\frac{1}{2P^0} f(P,x,s) 
  \langle P^s|\bar\psi\gamma^5\psi|P^s\rangle\, .
\end{equation}
If $f(P,x,s) = f(P,x)$ is spin-independent and the same for particles and antiparticles, and working in the comoving momentum $q$ as defined in~\cite{Ma:1995ey}, this gives
\begin{equation}\label{eq3:eom-pseudoscalar-tau}
  \ddot\phi + 3H\dot\phi +  m_\phi^2\phi =  
  -2g^2\phi\, \mathfrak{g}\, a^{-2}
  \int 4\pi \dd q \, q^2 \, \frac{1}{\sqrt{q^2 + a^2(m_0^2 + g^2\phi^2)}} f_0(q) \, ,
\end{equation}
where $f_0(q)$ is the Fermi-Dirac distribution,
\begin{equation}\label{eq3:f0}
    f_0(q) = \frac{1}{(2\pi)^3} \frac{1}{e^{q/T_0^\nu}+1}\, ,
\end{equation}$T^\nu_0$ the neutrino temperature today and $\mathfrak{g}$ the number of fermionic internal degrees of freedom: particles and antiparticles, and both spins. 

Here, there are three characteristic timescales,
\begin{itemize}
  \item the cosmological time, $H^{-1}$, which controls the Hubble friction term and the rate at which $f_0 = f_0(q)$ changes ($\frac{\dd f_0}{\dd t} = H\frac{\dd f_0}{\dd\log q}$). 
  \item the characteristic oscillation time of $\phi$. Naively, given by $m_\phi$, but might be modified by the interactions. 
  \item the timescale of neutrino dynamics, controlled by $E_p = \sqrt{\vec{p}^2+m_0^2+g^2\phi^2}$.
\end{itemize}
The interplay of these three scales defines the necessary physical assumptions. In the one hand, the energy of neutrinos (whether it is kinetic or the rest mass) is always $E_p\gtrsim T_{\nu,0}\sim 10^{-4}\, \rm eV$. Even after neutrino feedback, for $m_\phi\lesssim 10^{-17}\, \rm eV$, we always have $E_p\gg m_\phi$. Consequently, fermion dynamics are much faster than the oscillation of $\phi$. Taking $\phi$ to be constant during fermion dynamics, as done in the previous subsection, is valid. The same argument holds for electroweak interactions, with timescales $\, \rm GeV^{-1}$. Furthermore, the homogeneity of the early Universe allows us to neglect any spatial variations of $\phi$.  

On the other hand, ULDM is obtained in the regime $m_\phi\gg H$, such that the field oscillates in scales much shorter than the cosmological expansion. In this limit, oscillations of $\phi$ are averaged-out, and thus $\phi$ can be understood as a collection of CDM particles~\cite{OHare:2024nmr}.
However, in order to follow the dynamics of the interaction between the field and the fermions, e.g., see if fermions backreact in the dynamics of the field; an adiabatic approximation is required to rigorously average the oscillations out~\cite{Landau:1976mec}. 

\subsection{Adiabatic approximation}\label{sec3:pseudo-adiabatic}
In the framework of the adiabatic approximation, we define two different time variables. First, the slow cosmological timescale $H^{-1}$ is parametrised by the scale factor $a$. Second, the timescale of the field oscillation is parametrised with a proper time $t$. During an oscillation of $\phi$ (of timescale $t$), the scale factor $a$ can be held constant, i.e. $a'=0$. Thus, the equation of motion for $\phi$ becomes
\begin{equation}\label{eq3:eom-pseudoscalar-adiabatic}
  \ddot\phi + m_\phi^2\phi = 
  -2g^2\phi\, \mathfrak{g}\, 
  a^{-2}\int 4\pi \dd q\, q^2 \, \frac{1}{\sqrt{q^2 + a^2(m^2 + g^2\phi^2)}} f_0(q) \, ,
\end{equation}
This is the equation of a one-dimensional non-relativistic particle moving on a potential well. The adiabatic approximation assumes that the real motion of $\phi$ is well described by assuming $a$ constant in a complete period of oscillation $t_\phi$. 

In the adiabatic approximation, every observable $\mathcal{O}$ at cosmological timescales is replaced by its value averaged over the oscillation,
\begin{equation}\label{eq3:averaging-oscillation}
  \langle\mathcal{O}\rangle = 
  \frac{1}{t_\phi}\int_0^{t_\phi}\dd t\, \mathcal{O}(\phi(t),\dot\phi(t)) =
  \frac{1}{t_\phi}\oint \dd \phi\,  \frac{\mathcal{O}(\phi,\dot\phi(\phi))}{\dot\phi(\phi)}\, ,
\end{equation}
where $\oint$ marks an integral over an oscillation of $\phi$, of period
\begin{equation}
  t_\phi = \int_0^{t_{\phi}}\dd t = \oint \frac{\dd \phi}{\dot\phi(\phi)}\, .
\end{equation}
The period of oscillation would then define an oscillation frequency which we can interpret as an effective mass, $M_\phi = 2\pi/t_\phi$.
\begin{figure}[t]
  \centering
  \includegraphics[width = \linewidth, alt={plain-text This plot shows two subplots on top of each other, sharing the same x axis: the scale factor $a$ of the Universe in a logscale from 0.1 to 1. In the upper plot, the y axis shows the value of the field $\phi$, normalized by its amplitude today $A_0$. The value of the field oscillates around 0, and the amplitude of the oscillations increases to the past (to the left). In the lower plot, the y axis shows the value of the neutrino effective mass $m_\nu$. Since the field oscillates in time, the mass also oscillates from $m_0$ to $\sqrt{m_0+g^2A^2}$. A red line then is drawn in front of the oscillatory pattern, to show the averaged value $\langle m_\nu\rangle$, which increases smoothly to the past.}]{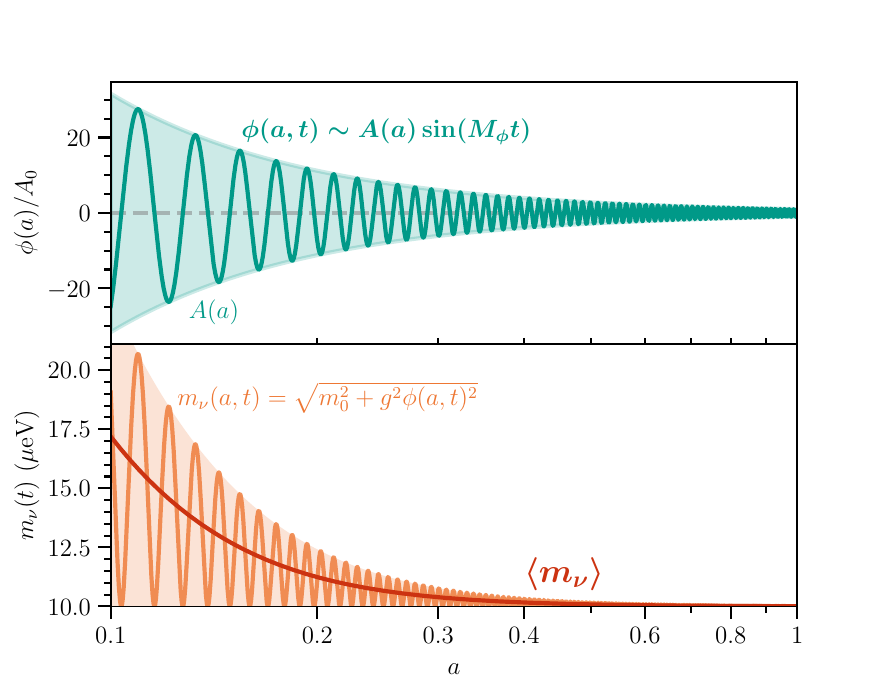}
  \caption{Illustration of the adiabatic approximation method. The top plot shows how the ULDM field $\phi(a,t)$ is oscillating rapidly around zero with amplitude $A(a)$ and frequency $M_\phi$. As a consequence of the oscillating field, the bottom plot shows how the mass of the neutrino, $m_\nu(a,t)$, also oscillates rapidly with time. The adiabatic approximation allows to compute the averaged neutrino mass, $\langle m_\nu\rangle$, or the average of any other observable $\mathcal{O}$ as in \cref{eq3:averaging-oscillation}. Curves here correspond to $m_0 = 10\, \mu\mathrm{eV}$, $g = 1.5\times 10^{-20}$ and $m_\phi = 10^{-19}\, \mathrm{eV}$, and $A_0 \equiv A(a=1)$. For illustration purposes, the shown oscillation frequency is reduced by $\mathcal{O}(10^{21})$ compared to $M_\phi$.}
  \label{fig3:adiabatic-approximation}
\end{figure}
The total energy density is given by
\begin{equation}\label{eq3:adiabatic-rho}
  \rho(a,t) = \frac{1}{2}\dot\phi^2 - \frac{1}{2}m_\phi^2 \phi^2 + \rho_\nu(a,\phi)\, .
\end{equation}
This is similar to the standard ULDM scenario, but with an additional term accounting for the energy in the fermion sector,
\begin{equation}\label{eq3:fermion-energy-density}
  \rho_\nu(a,\phi) = a^{-4}\mathfrak{g}\int 4\pi \dd q\, q^2 \sqrt{q^2+a^2(m_0^2+g^2\phi^2)}\, f_0(q)\, .
\end{equation}
Within an oscillation we can treat the total energy as a constant of motion, $\rho(a,t) \simeq \langle\rho(a,t)\rangle \equiv \rho(a)$. 
This allows us to use~\cref{eq3:adiabatic-rho} to find a closed expression for $\dot\phi$,
\begin{equation}
  \dot\phi = \sqrt{2}\sqrt{\rho(a) - m_\phi^2\phi^2/2 - \rho_\nu(\phi) }
\end{equation}
The adiabatic approximation formalism leads to the result that
\begin{equation}\label{eq3:adiabatic-invariant}
  \mathcal{I}(a) \equiv \frac{1}{\sqrt{2}} a^3t_\phi\langle \dot\phi^2\rangle = 
  a^3\oint\dd \phi\, \sqrt{\rho(a) - \frac{1}{2}m_\phi^2\phi^2 - \rho_\nu(\phi)}\, ,
\end{equation}
is an invariant quantity for the whole cosmological evolution, i.e., $\dd\mathcal{I}/\dd a = 0$. This adiabatic invariant allows us to compute the total energy density $\rho(a)$ of the system as a function of the scale factor, given some initial conditions. The derivation of this result is presented in \cref{app:formalism}.

Instead of $\rho(a)$, we work in terms of the amplitude of oscillations of the pseudoscalar, $A(a)$. In particular, $\phi=A$ is reached when $\dot\phi=0$, and thus
\begin{equation}\label{eq3:rho-amplitude}
  \rho(a) = \frac{1}{2}m_\phi^2 A(a)^2 + \rho_\nu(a,A(a))\, .
\end{equation}
Then,~\cref{eq3:adiabatic-invariant} allows to compute the evolution of $A(a)$. In the $g\to 0$ limit, the adiabatic invariant becomes
\begin{equation}
  \mathcal{I}(a) = a^3\oint \dd\phi\,  \sqrt{\frac{1}{2}m_\phi^2(A^2-\phi^2)}  =
  \frac{\pi}{2}m_\phi\, a^3A^2 \, , 
\end{equation}
which means that $A(a) \sim a^{-3/2}$. As expected, in the $g\to 0$ limit $\phi$ behaves as CDM, with $\rho(a)\sim a^{-3}$. In the limit where the interaction potential dominates, i.e., $m_\phi^2A^2 \ll \rho_\nu(a,A(a))$, the adiabatic invariant becomes
\begin{equation}
\begin{split}
  \mathcal{I}(a) = a \oint \dd\phi\, \sqrt{\,\mathfrak{g}\int 4\pi q^2\, \dd q
  \left(\sqrt{q^2+g^2a^2A^2}-\sqrt{q^2+g^2a^2\phi^2}\right)\, f_0(q)}\, .
\end{split}
\end{equation}
By changing variables to $\tilde\phi = a\phi$, one can check that $\mathcal{I}(a,A) = \mathcal{I}(aA)$, thus $aA=\mathrm{constant}$. In the limit where the interaction potential dominates, the pseudoscalar field does not scale like CDM, but following $A(a) \sim a^{-1}$. This is due to the neutrino-dominated potential having a different scaling dependence to the $m_\phi$ potential. Let us look into this further.

\section{Cosmological evolution}\label{sec3:pseudo-phenomenology}
Now that the adiabatic approximation fully follows the bidirectional feedback between the neutrino sector and the ULDM field, in this Section we compute important cosmological variables affected by the interaction. 
All the shown results are computed in a modified version of \texttt{CLASS}~\cite{Lesgourgues:2011re, Blas:2011rf, Lesgourgues:2011rg, Lesgourgues:2011rh} which implements the full adiabatic approximation method.

\subsection{Energy density and effective mass}
\begin{figure}[t]
  \centering
  \includegraphics[width = \linewidth, alt={plain-text this plots shows four subplots, with the same axis in each of them, but evaluated at different times. In each plot, the x axis shows the field $\phi$ normalized by the amplitude $A$ (from -1 to 1), and the y axis shows the potential energy, in arbitrary units. In each plot, a green line shows the bare mass $m_\phi$ potential, and an orange line the potential due to neutrino interactions. In the upper left subplot, the quadratic $m_\phi$ potential dominates and scales back in time with $a^{-3}$. Since the neutrino potential scales with $a^{-4}$ is the upper right plot shows the time where both potentials match, defined as $a_{\mathrm{tr}}$. Then, in the lower row, where $a < a_{\mathrm{tr}}$ the total potential energy is dominated by neutrinos. In the lower left, the full neutrino potential is compared to the quadratic potential $\delta m_\phi^2\phi^2$, and look very similar with a small difference at large amplitudes. In the lower right, the potential if neutrinos became non-relativistic. At large amplitudes, it becomes linear.}]{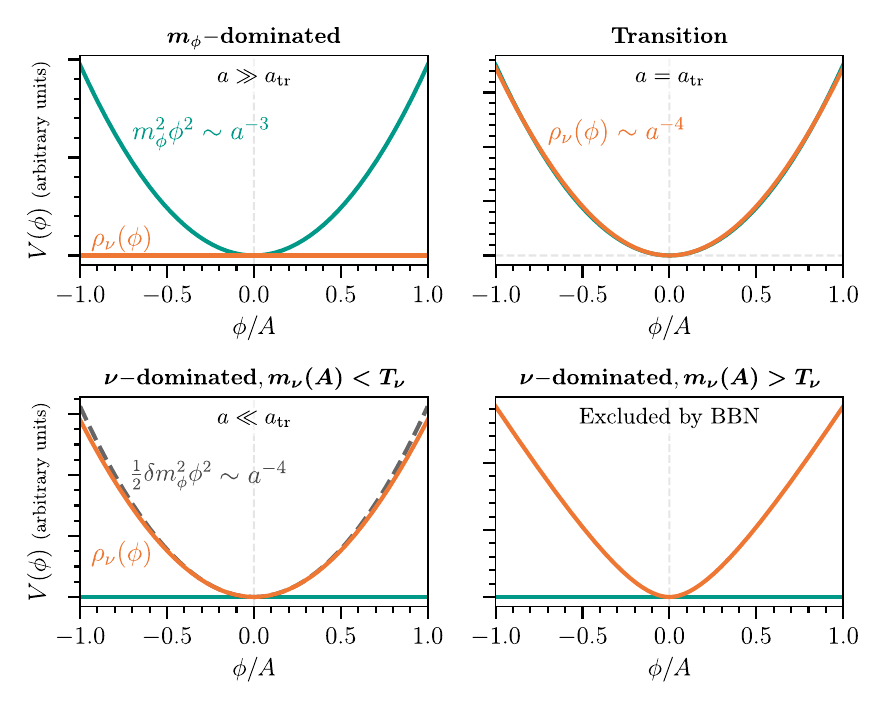}
  \caption{Evolution of the potential as in \cref{eq3:pseudo-potential}. Left to right, top to bottom goes back in time. Today, the effect of neutrino interactions is negligible, and $V(\phi)=m_\phi^2\phi^2/2$. However, $\rho_\nu(\phi)$ scales as $a^{-4}$, and therefore at some point ($a=a_{\mathrm{tr}}$) in the past the neutrino potential equals the bare-$m_\phi$ potential. From that time on, $V(\phi)\sim a^{-4}$ and the scaling of $A(a)$ changes. If the momentum of neutrinos is larger than their effective mass, the $V(\phi)$ is quadratic, but if it becomes comparable, $V(\phi)$ becomes linear at large $\phi$.} 
  \label{fig3:potential-energy-evolution} 
\end{figure}

From the energy-momentum tensor, 
\begin{equation}
  \rho(a) = \langle\rho(a,t)\rangle = 
  \frac{1}{2}\langle\dot\phi^2\rangle + \frac{1}{2}m_\phi^2\langle\phi^2\rangle + 
  \langle\rho_\nu(a,\phi)\rangle \equiv \frac{1}{2}\langle\dot\phi^2\rangle + 
  \langle V(\phi)\rangle \, ,
\end{equation}
which coincides with \cref{eq3:rho-amplitude}, and the pseudoscalar potential is
\begin{equation}\label{eq3:pseudo-potential}
  V(\phi) = \frac{1}{2}m_\phi^2\phi^2 + \rho_\nu(a,\phi)\, .
\end{equation}
The shape and scaling of this potential at different epochs are shown in~\cref{fig3:potential-energy-evolution}.  
Firstly, $m_\phi^2\phi^2$ is a quadratic potential, which scales as $a^{-3}$ if $\phi$ behaves as standard CDM. Secondly, $\rho_\nu\sim a^{-4}$. Then, at some point in the past, defined as $a_{\mathrm{tr}}$ in \cref{fig3:potential-energy-evolution}, $\rho_\nu$ dominates and $V(\phi)$ changes scaling from $a^{-3}$ to $a^{-4}$, while still being quadratic. 
Finally, if $gA \gg |\vec{p}|$, neutrinos can become non-relativistic, and then $V(\phi)$ becomes linear at large $\phi$. The couplings required for this scenario are ruled out from the bounds obtained in \cref{sec3:pseudo-results}.

In a first approximation, we can assume that the linear momentum of neutrinos is much larger than their instantaneous mass, i.e., $T_\nu \gg gA$. 
We will refer to this regime as the linearised regime. In this limit, we can expand $\rho_\phi$ up to first order in $\phi$ and get
\begin{equation}
  \rho(a,t) = \frac{1}{2}\dot\phi^2+ \frac{1}{2}m_\phi^2\phi^2 + \frac{1}{2} \delta m_\phi^2 \phi^2 + \rho_\nu(a,0)\, ,
\end{equation}
where
\begin{equation}\label{eq3:delta_mphi2}
  \delta m_\phi^2 = 2\, \frac{\partial \rho_\nu(a,\phi)}{\partial \phi^2}\biggr|_{\phi = 0} = 
  a^{-2}\, g^2\, \mathfrak{g}\int 4\pi \dd q\, q^2\frac{1}{\sqrt{q^2+a^2m_0^2}}f(q)\, 
\end{equation}
is an effective pseudoscalar mass from the energy expense necessary to provide neutrinos their mass. This effective mass scales as $\delta m_\phi^2\sim a^{-2}$, as follows. The energy of a single neutrino coupled to $\phi$, when $T_\nu \gg gaA\gg m_0$, is
\begin{equation}
  E_p = \sqrt{\vec{p}^2+ m_0^2+ g^2\phi^2} \simeq |\vec{p}| + \frac{g^2\phi^2}{2|\vec{p}|}\, ,
\end{equation}
which implies an increment in energy $\sim g^2\phi^2/|\vec{p}|$. For fixed $\phi$, this scales as $\sim a$, but the number density of neutrinos which receive the mass scales with $a^{-3}$. Thus, $\delta m_\phi^2 \sim a^{-2}$. 

\begin{figure}[t]
  \centering
  \includegraphics[width = 0.9\linewidth, alt={plain-text this plots shows the evolution of the energy densition, normalized by $a^3/\rho_{crit,0}$ as a function of the scale factor, from 1e-7 to today. Today, the total energy density equals the phi energy density and is constant (i.e. scales with $a^{-3}$), while rho_nu falls as $a^{-1}$ (i.e. $a^{-4}$). When $a \sim a_{\mathrm{tr}}\sim 10^{-5}$, then all the energy densities increase to the past linearly (i.e. $a^{-4}$). In the decoupled scenario, the CDM energy density remains always increasing with $a^{-3}$}]{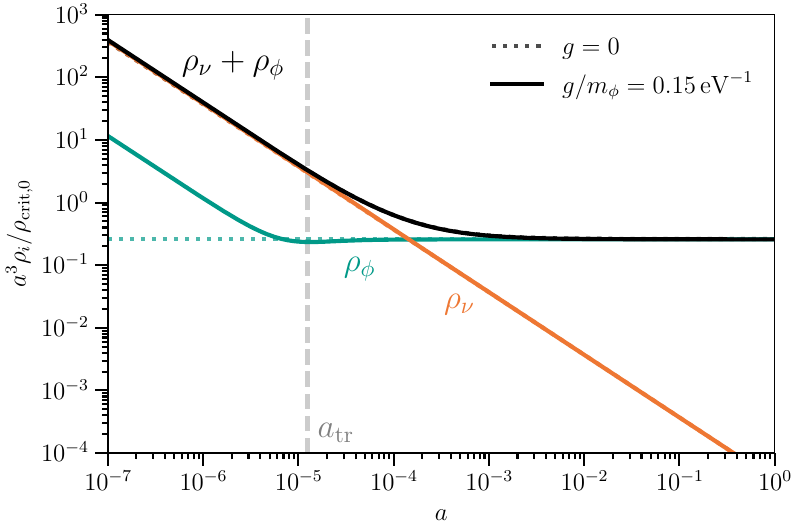}
  \caption{Evolution of the energy density of the $\nu+$ULDM fluid, in the interacting scenario (solid) and in the decoupled scenario (dotted). In both cases, $m_0 = 10^{-5}\, \mathrm{eV}$. A vertical dashed line shows the time of transition between $m_\phi$-domination and $\nu$-domination. $\rho_\phi$ (teal) and $\rho_\nu$ (orange) can only be understood as independent fluids for $a\gg a_{\mathrm{tr}}$. For $a < a_{\mathrm{tr}}$, their separation is artificial.}
  \label{fig3:pseudo-energy-density}
\end{figure}

Then, the total effective mass of the pseudoscalar field is given by
\begin{equation}\label{eq3:linearised-effective-mass}
  M_\phi^2 = 2\, \frac{\partial^2V(\phi)}{\partial\phi^2}\biggr|_{\phi=0} = m_\phi^2 + \delta m_\phi^2 \sim \begin{cases}
    m_\phi^2 \sim \mathrm{constant} & a > a_{\mathrm{tr}} \\
    \delta m_\phi^2 \sim a^{-2} & a < a_{\mathrm{tr}}
  \end{cases} \, ,
\end{equation}
where $a_{\mathrm{tr}}$ is quantitatively defined through $m_\phi^2 = \delta m_\phi^2$,
\begin{equation}\label{eq3:pseudo-transition}
  a_{\mathrm{tr}}^2 = \frac{g^2}{m_\phi^2}\, \mathfrak{g}\int 4\pi \dd q\, q^2\frac{1}{\sqrt{q^2+a^2m_0^2}}f(q)\sim \left(8.4\times 10^{-5}\, \frac{g/m_\phi}{\mathrm{eV}^{-1}}\right)^2\, ,
\end{equation}
where we have used $T_{\nu, 0} = 0.1681\, \mathrm{meV}$ and $\mathfrak{g} = 6$ for the numerical estimate.
Then, the equation of motion \cref*{eq3:eom-pseudoscalar-adiabatic} simplifies to
\begin{equation}
  \ddot \phi + M_\phi^2\phi = 0\, ,
\end{equation}
which is the equation of a quadratic harmonic oscillator, with standard solution
\begin{equation}\label{eq3:linearised-solution}
  \phi(a,t) = A(a) \sin(M_\phi t+\varphi_0)\, ,
\end{equation}
with $\varphi_0$ an initial phase. In this case, it is trivial to average over an oscillation of the field, and for instance the total energy density is given by
\begin{equation}\label{eq3:linearised-rho}
  \rho(a) = \frac{1}{2}M_\phi^2 A^2 + \rho_\nu(a,0)\, , 
\end{equation}
where $\rho_\nu(a,0)$ is the energy density of neutrinos as if they were decoupled. The evolution of this energy density is compared to standard $\nu+$ULDM in \cref{fig3:pseudo-energy-density}. While everything looks very similar, we can see that $\rho_\phi \equiv \frac{1}{2}\langle\dot\phi^2\rangle + \frac{1}{2}m_\phi^2\langle\phi^2\rangle$ scales like radiation for $a < a_{\mathrm{tr}}$.

The radiation behaviour of $\rho_\phi$ can be understood at the same time as one explains why $a_{\mathrm{tr}}$ also describes the transition between $A\sim a^{-3/2}$ and $A\sim a^{-1}$ from~\cref{eq3:adiabatic-invariant}. This stems from the conservation of the number of $\phi$ particles, i.e., the scaling of its number density as $n_\phi\sim a^{-3}$~\cite{Huang:2021kam}. Its energy density is, in turn, $\rho_\phi = M_\phi n_\phi$. For $a > a_{\mathrm{tr}}$, $M_\phi$ is constant, and thus $\rho_\phi = m_\phi n_\phi\sim a^{-3}$. Since $\rho_\phi \equiv \frac{1}{2}m_\phi^2A^2$, this means that $A\sim a^{-3/2}$, as standard CDM. However, when $a< a_{\mathrm{tr}}$, $\rho_\phi = \delta m_\phi n_\phi\sim a^{-4}$, as if it were radiation. Since $\rho_\phi \equiv \frac{1}{2}\delta m_\phi^2A^2$, this means that $A\sim a^{-1}$. Since $\phi$ is homogeneous, we expect perturbations to still behave as CDM with zero momentum, but the exact treatment of perturbations is left for future work. In order to better understand the effect of this transition in the expansion history of the Universe, we now quantitatively compute the equation of state.

\subsection{Pressure and equation of state}\label{sec3:pseudo-pressure-eos}
From the energy-momentum tensor, the pressure of the fermion-pseudoscalar fluid is
\begin{equation}
  p(a) =  
  \frac{1}{2}\langle\dot\phi^2\rangle - \frac{1}{2}m_\phi^2\langle\phi^2\rangle + 
  \langle p_\nu(a,\phi)\rangle\, ,
\end{equation}with
\begin{equation}
  p_\nu(a,\phi) = \mathfrak{g}\int \dd^3 \vec{p}
  \frac{p^2}{\sqrt{p^2+m_0^2+g^2\phi^2}}\, f_0(pa) = 
  a^{-4}\mathfrak{g}\int \frac{4\pi\dd q\, q^4}{\sqrt{q^2+a^2(m_0^2+g^2\phi^2)}}f_0(q)\, . 
\end{equation}
In the linearised regime approximation, this can be simplified to 
\begin{equation}
  p(a,t) = \frac{1}{2}\dot\phi^2 - \frac{1}{2}m_\phi^2\phi^2 - \frac{1}{2} \delta\mu_\phi^2 \phi^2 + p_\nu(a,0)\, ,
\end{equation}
where
\begin{equation}
  \delta\mu_\phi^2 = -2\,\frac{\partial p_\nu(a,\phi)}{\partial\phi^2}\biggr|_{\phi=0} = 
  a^{-2}g^2\mathfrak{g}\int 4\pi \dd q\, \frac{q^4}{(q^2+a^2m_0^2)^{3/2}}f_0(q)
\end{equation}
is a higher-order integral of the momentum distribution. Now, plugging the solution from \cref{eq3:linearised-solution} and averaging over an oscillation, 
\begin{equation}\label{eq3:linearised-pressure}
  p(a) = \frac{1}{4}\left(\delta m_\phi^2-\delta \mu_\phi^2\right)A^2 + p_\nu(a,0)\, . 
\end{equation}
Since $\delta m_\phi^2\ge \delta\mu_\phi^2$, the scalar part of the pressure is never negative.
Then, one can define an equation of state for the total fluid,
\begin{equation}
  w(a) = \frac{p(a)}{\rho(a)} = 
  \frac{\left(\delta m_\phi^2-\delta \mu_\phi^2\right)A^2/4 + p_\nu(a,0)}{(m_\phi^2 +\delta m_\phi^2)A^2/2 + \rho_\nu(a,0)}\, .
\end{equation}

\begin{figure}[t]
  \centering
  \includegraphics[width = 0.9\linewidth, alt={plain-text this plots shows the evolution of the equation of state $w = p/\rho$ as a function of the scale factor, from 1e-7 to today. Today, $w = 0$, while at 1e-7, $w = 1/3$. Around 1e-4, the equation smoothly transitions from one value to the other. Different values of the coupling are then shown, from $g/m_\phi = 0.15/eV$ to $g/m_\phi = 1.5/eV$. The larger the coupling, the more delayed the transition from radiation to matter is.}]{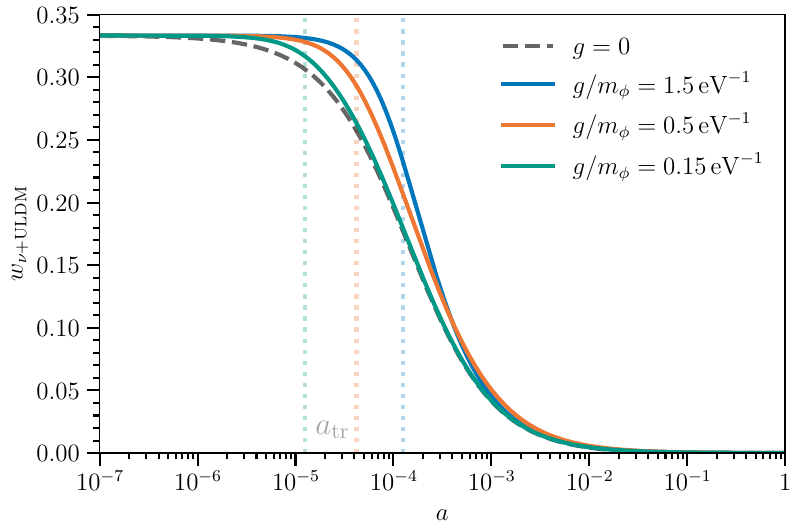}
  \caption{Evolution of the equation state of the $\nu+$ULDM fluid for interacting scenarios (solid lines) and a decoupled scenario (dotted). Vertical dotted lines shows $a_{\mathrm{tr}}$ (transition between $\nu$-domination and $m_\phi$ domination) for the different couplings, as in \cref{eq3:pseudo-transition}. \textit{The coupled fluid stays relativistic for a longer period.}}    
  \label{fig3:pseudo-equation-of-state}
\end{figure}
Now, we can simplify this expression further. In the early universe, for allowed values of $m_0$, $T_\nu \gg m_0$, and $m_0$ can be neglected in integrals over the distribution function. In this limit, $\delta m_\phi^2 = \delta\mu_\phi^2$, and therefore
\begin{equation}
  p(a) = p_\nu(a,0) = \frac{1}{3}\rho_\nu(a,0)\, ,
\end{equation}
the pressure of the total fluid is only the pressure of the neutrino part, \textit{as if they were uncoupled from the pseudoscalar field}, which are completely relativistic at early times. This simplifies the equation of state to
\begin{equation}\label{eq3:linear-eos-pseudo}
  w(a) = \frac{1}{3}\frac{\rho_\nu(a,0)}{\rho_\nu(a,0) + (m_\phi^2 + \delta m_\phi^2)A^2/2} = \frac{1}{3}\left[1- \left(1+\frac{\rho_\nu(a,0)}{(m_\phi^2 + \delta m_\phi^2)A^2/2}\right)^{-1}\right]\, .
\end{equation}
\Cref{fig3:pseudo-equation-of-state} shows the evolution of $w(a)$ for different couplings. For the standard scenario $g\to 0$, $\delta m_\phi^2 = 0$. Then, if $m_\phi^2 A^2 \gg \rho_\nu(a,0)$, $w = 0$ (CDM domination); and if $m_\phi^2 A^2 \ll \rho_\nu(a,0)$, $w = 1/3$ (relativistic neutrino domination). In the coupled scenario, the modified scaling of $A(a)$ makes $w(a)$ grows faster to $1/3$ than in the standard case. One of the consequences of the fluid staying relativistic for a longer time is that the time of equality between matter and radiation is delayed~\cite{Huang:2021kam}. However, for couplings allowed by BBN this is a subleading effect, as we will explore below.

\subsection{Initial conditions}\label{sec3:pseudo-initial-conditions}
We want to impose that $\phi$ makes up all DM, namely $\Omega_{\phi,0}h^2 =\Omega_{c,0}h^2 = 0.1200$~\cite{Planck:2018vyg}, with $h = 0.6766$ the reduced Hubble constant. This requires that the amplitude of the field today is
\begin{equation}\label{eq3:amplitude-DM-abundance}
  A(a=1) \equiv A_0 = \frac{\sqrt{2\Omega_{c,0}\, \rho_{\mathrm{crit}}}}{m_\phi}\, ,
\end{equation}
assuming $m_\phi > \delta m_\phi$ today. 
The value of $A$ today fixes the value of the adiabatic invariant $\mathcal{I}(a) = \mathcal{I}_0$, and allows to compute $A(a)$ at all times if we know $f_0(q)$. As proved in \cref{app:formalism}, the coupling with $\phi$ does not change the evolution of $f_0(q)$, and therefore its shape is frozen from its initial conditions, i.e., neutrino decoupling. 

\begin{figure}[t]
  \centering
  \includegraphics[width = \linewidth, alt={plain-text this plots shows two subplots on top of each other, with the scale factor in the x axis from 1e-12 to today. In the upper plot, the amplitude field is shown, normalized by $am_\phi$, such that all curves collapse in the same value today. As one goes to the past, the scaling of the amplitude changes from $a^{-3/2}$ to $a^{-1}$, and when misalignment happens, the amplitude becomes constant. Three curves are shown with different coupling and $m_\phi$. The smaller $g/m_\phi$, the earlier the transition and the larger the amplitude in the early universe. In the lower plot, the effective mass of the field $M_\phi$ is compared to the Hubble rate, $3H$. Models with the same $m_\phi$ have the same $M_\phi$ today, but models with the same coupling have the same $M_\phi$ before $a_{\mathrm{tr}}$. Vertical dashed lines show the misalignment time where $M_\phi = 3H$, which mostly depends on $g$.}]{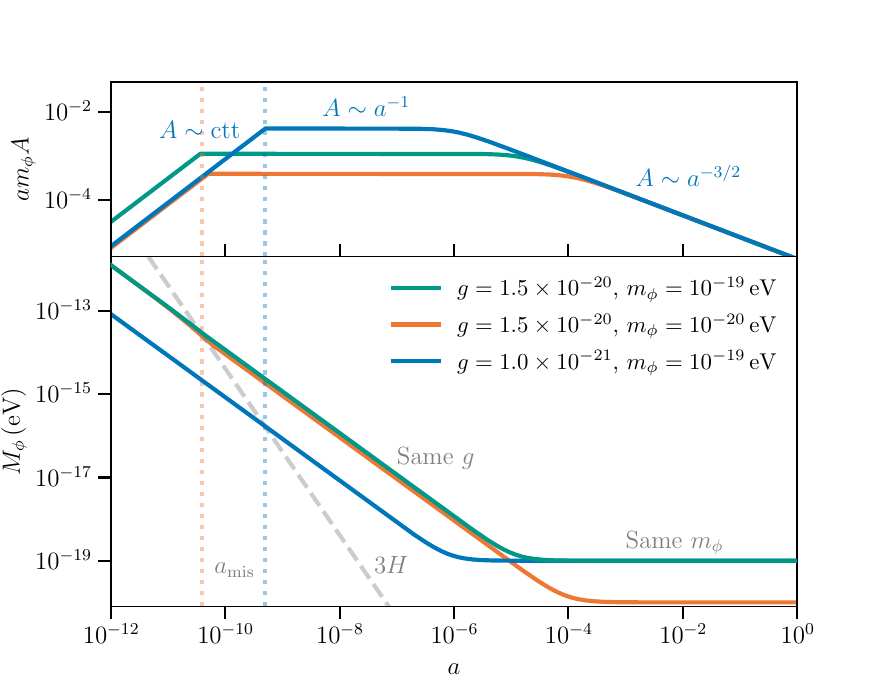}
  \caption{Effective mass $M_\phi$, misalignment mechanism and evolution of the amplitude. From left to right, initially $3H>M_\phi$, and thus $A(a)$ remains constant. At misalignment, $3H\sim M_\phi$, the field starts evolving as $a^{-1}$ if $a < a_{\mathrm{tr}}$ or as $a^{-3/2}$ if $a> a_{\mathrm{tr}}$. Curves with the same coupling $g$ satisfy the misalignment condition at similar times, while fields with the same $m_\phi$ arrive at the same point in $M_\phi$. Here, the amplitude today is set to match the abundance of DM, and thus $A_0\sim m_\phi^{-1}$.}
  \label{fig3:evolution-amplitude-misalignment}
\end{figure}

The early-Universe physics and, in particular, decoupling depend on the time of the misalignment mechanism. In standard ULDM, oscillations start when the potential energy term is larger than the Hubble friction, i.e., $m_\phi > 3H$. However, the interaction modifies the condition to $M_\phi > 3H$. This advances misalignment, $a_{\mathrm{mis}}$, to earlier times, as shown in \cref{fig3:evolution-amplitude-misalignment}. 

Since $A_0\propto m_\phi^{-1}$ and $A$ always appears as $gA$, many of the cosmological observables depend only on $g/m_\phi$. However, $\delta m_\phi^2\sim g^2$ is independent of $m_\phi$. Thus, the misalignment mechanism $3H<M_\phi$ breaks the degeneracy between $g$ and $m_\phi$. We find that, for couplings such that $\delta m_\phi^2>m_\phi^2$, misalignment happens at 
\begin{equation}
  a_{\mathrm{mis}} \simeq  10^{-11}\left(\frac{5\times 10^{-20}}{g}\right)\, .
\end{equation}  
Decoupling happens approximately at $T_\gamma \sim 2\, \mathrm{MeV}$, which --assuming a naive linear scaling for temperature-- corresponds to $a\sim 10^{-11}$. That is, misalignment happens after decoupling roughly if $g \lesssim 5\times 10^{-20}$. 
Two scenarios open up,
\begin{itemize}
  \setlength\itemsep{-2pt}
  \item Before misalignment, $m_\nu$ is constant, $T_\nu/m_\nu$ grows to the past and neutrinos are relativistic soon before misalignment. If misalignment happens after decoupling, $g \lesssim 5\times 10^{-20}$, then neutrinos are relativistic at decoupling and effects from the interaction are negligible. If they are, their distribution function at later times will be a relativistic Fermi-Dirac such as~\cref{eq3:f0}.
  \item If misalignment happens while neutrinos are in thermal equilibrium with the baryon plasma, $g \gtrsim 5\times 10^{-20}$, then the growth of $A(a)$ might make them non-relativistic. Then, their distribution function will deviate from a Fermi-Dirac with $m_\nu/T_\nu$ corrections. The full treatment requires thermal masses~\cite{Venzor:2020ova}, but also to consider the effect of scatterings mediated by $\phi$, far from the extent of this work. 
\end{itemize}

With this in mind, we assume that misalignment happens after decoupling, and we approximate the misalignment to be instantaneous at $3H = M_\phi$. Before misalignment, we set $\phi = A$ and no oscillation is needed. At misalignment, $\phi$ starts to oscillate with $M_\phi$ and all observables are averaged as in~\cref{eq3:averaging-oscillation}. This approximation produces an artificial discontinuity between unaveraged and averaged variables. A smooth misalignment should not strongly modify the results.

\subsection{Massive neutrinos and $N_{\mathrm{eff}}$}\label{sec3:mnu-neff}
The neutrino-feedback into the pseudoscalar field has two main phenomenological consequences, an increased effective mass for neutrinos and an increase of early radiation. 

Firstly, neutrinos will acquire a mass given by
\begin{equation}
  \langle m_\nu \rangle = \left\langle m_\nu(\phi)\right\rangle = \left\langle \sqrt{m_0^2 + g^2\phi^2} \right\rangle\, ,
\end{equation}
where the average is taken over the oscillation of the field. In the linearized regime, and for early enough in the Universe, $\langle m_\nu \rangle \simeq gA/\sqrt{2}$. 
\begin{figure}[t]
  \centering
  \includegraphics[width = \linewidth, alt={plain-text this plot has two subplots on top of each other, sharing the same x axis: the scale factor between 1e-12 and today. In the upper plot, the averaged effective mass of the neutrino, $\langle m_\nu \rangle$ is shown. Today the mass is constant and $m_0$, but soon early enough it increases with $a^{-3/2}$, and for $a<a_{\mathrm{tr}}$ it begins to increase linearly $a^{-1}$, and finally for $a<a_{\mathrm{mis}}$ it becomes constant. This is shown for three increasing $g/m_\phi$, from 0.01/eV to 1.5/eV. The larger $m_\phi$, the larger the mass. A gray dashed region shows the region where $m_\nu > T_\nu$, i.e., where neutrinos become non-relativistic. The green line, with $g/m_\phi = 0.15$ is just slightly below the region. In the lower plot, $\langle m_\nu\rangle$ is normalized to $T_\nu\propto a^{-1}$ for better visualization of the different regions.}]{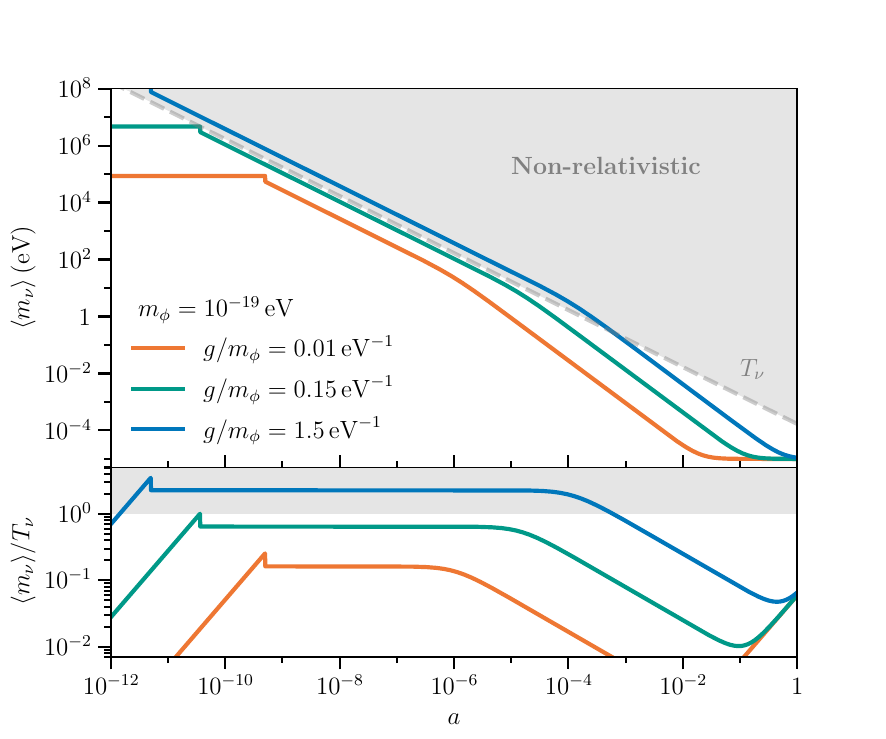}
  \caption{Evolution of average neutrino mass, $\langle m_\nu \rangle$, and its \textit{non-relativisticness}, i.e., $\langle m_\nu \rangle/T_\nu$. Initially, $m_\nu\sim m_0 = 10^{-5}\, \mathrm{eV}$, and soon $\langle m_\nu\rangle \sim gA/\sqrt{2}$. Thus, $\langle m_\nu\rangle$ follows $A(a)$, shown in \cref{fig3:evolution-amplitude-misalignment}. For large (excluded) couplings, neutrinos become non-relativistic. Since the linearised approximation is an expansion on $\left(\left\langle m_\nu\right\rangle/T_\nu\right)^2$, the bottom plot justifies the validity of the approximation for non-excluded couplings.}
  \label{fig3:pseudo-neutrino-mass}
\end{figure}
\Cref{fig3:pseudo-neutrino-mass} compares the evolution of the neutrino mass and $T_\nu$, which is related to the mean momentum of the distribution. While early enough, the neutrino mass can reach $\mathcal{O}(1)\, \mathrm{MeV}$, the important quantity is the \textit{relativisticness} (i.e., velocity) of neutrinos, $T_\nu/\left\langle m_\nu\right\rangle$. That is, the interaction with the field can make them non-relativistic, $\left\langle m_\nu\right\rangle \gg T_\nu$. \Cref{fig3:pseudo-neutrino-mass} also shows how $T_\nu/\left\langle m_\nu\right\rangle$ diminishes as $a^{-1/2}$ for $a > a_{\mathrm{tr}}$, while  $T_\nu/\left\langle m_\nu\right\rangle$ becomes constant for $a < a_{\mathrm{tr}}$. At misalignment, $\phi$ freezes and neutrinos become relativistic soon enough.

\begin{figure}[t]
  \centering
  \includegraphics[width = 0.9\linewidth, alt={plain-text in the x axis, the scale factor from 1e-12 to today; in the y axis, the contribution to $N_{eff}$ of the model. A data point is shown at the CMB epoch ($a\sim 10^{-3}$) and at BBN ($a\sim 10^{-7}$), both of them quite compatible with zero. Three curves shows the extra degrees of freedom that the model introduces, for increasing $g/m_\phi$ between 0.01/eV and 0.5/eV. The three curves are well compatible with the CMB data point, but increase in pre-CMB epoch, for $a<a_{tr}$. Then, $\Delta N_{eff}$ stays flat until misalignment, where it drops to negative values shortly and soon goes back to zero. The orange line, with $g/m_\phi = 0.15/eV$ is already at 2sigma from the BBN value, while $g/m_\phi =0.01/eV$ is quite close to zero (as in the decoupled case).}]{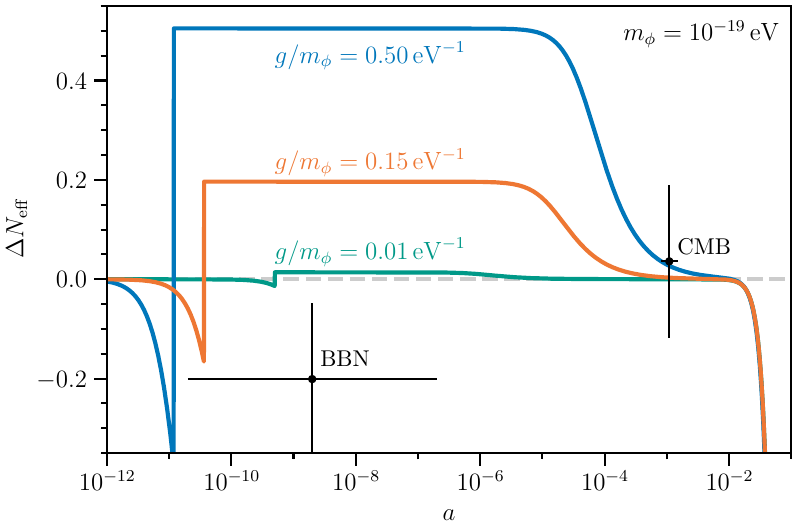}
  \caption{Number of relativistic neutrino species for different values of the coupling. The dilation of the radiation phase, shown in \cref{fig3:pseudo-equation-of-state}, increases the amount of radiation in the early universe. Data points show the measurements of $N_{\mathrm{eff}}$ at CMB and BBN, with horizontal errorbars showing the approximate duration of each epoch. Couplings which have a significant impact on BBN are negligible at CMB. \textit{BBN is the best epoch to constrain this model.}}
  \label{fig3:pseudo-neff}
\end{figure}
Secondly, if we fix the right DM abundance today, a longer radiation epoch as shown in~\cref{fig3:pseudo-equation-of-state} increases the amount of radiation in the early Universe, conventionally described by the number of relativistic degrees of freedom,
\begin{equation}\label{eq3:neff-definition}
  N_{\mathrm{eff}} = \frac{\rho_{\mathrm{rad}}-\rho_\gamma}{\frac{7}{8}\left(\frac{4}{11}\right)^{4/3}\rho_\gamma}\, ,
\end{equation}
where $\rho_{\mathrm{rad}},\, \rho_\gamma $ are the energy densities for all radiation and photons, respectively. 
We quantify the contribution of our model with $\Delta N_{\mathrm{eff}} = N_{\mathrm{eff}} - 3.044$. 
\Cref{fig3:pseudo-neff} shows the evolution of $\Delta N_{\mathrm{eff}}$ in our model. 
The radiation excess is specially significant at the pre-CMB epoch, as expected from \cref{eq3:pseudo-transition}. As a consequence, we expect BBN to put better constraints on the neutrino feedback of the model. 
\Cref{fig3:pseudo-neff} also shows a short depletion of $N_{\mathrm{eff}}$ at the time of misalignment, which is an artifact of the instantaneous misalignment approximation. Namely, since at $a<a_{\mathrm{mis}}$ we set $\phi = A$ and skip oscillations, then $m_\nu = gA$ instead of $\langle m_\nu \rangle = gA/\sqrt{2}$, neutrinos become slightly non-relativistic and decrease $\rho_{\mathrm{rad}}$. However, as soon as the plasma heats the constant mass of neutrinos becomes negligible, thus making them relativistic and retrieving $\Delta N_{\mathrm{eff}}=0$.

\section{Big Bang Nucleosynthesis}\label{sec3:pseudo-bbn}
Big Bang Nucleosynthesis (BBN) is the process by which the primordial light elements are formed in the first twenty minutes of the Universe, from the reactions between protons ($p$), neutrons ($n$), electrons, and electron neutrinos. The abundances of the primordial light elements are measured by astrophysical observations. Helium-4 and deuterium abundances are the best understood and have been measured to be~\cite{Pitrou:2018cgg,ParticleDataGroup:2024cfk}
\begin{equation}\label{eq3:bbn-observed-abundances}
    \begin{split}
    Y_\mathrm{P} \equiv X_{^4\mathrm{He}} &= 0.245\pm 0.003\, , \\
    \mathrm{D/H} \equiv n_{\mathrm{D}}/n_{\mathrm{^1H}} &= (2.547\pm 0.025)\times 10^{-5}\, , \\
    \end{split}
\end{equation}
respectively. A joint fit with the CMB provides an excellent agreement between the two cosmological probes, and the excellent agreement with the SM makes BBN a powerful probe of BSM physics in the very-early Universe.
In this Section, we implement our model in the full calculation of BBN abundances and constrain the model from there. 

The impact of the model in BBN is two-fold. First, $\Delta N_{\mathrm{eff}}$ will increase the Hubble rate, thus making nuclear processes less efficient and modifying the duration of nucleosynthesis. 
Secondly, the non-zero mass of neutrinos will modify the kinematics of the reactions and thus the interaction rates $\Gamma$. For instance, if neutrinos had mass $m_\nu > \Delta - m_e$, with $\Delta = m_n- m_p\simeq 1.29333\, \mathrm{MeV}$, neutron decay would be forbidden, greatly increasing $Y_{\mathrm{P}}$. 

\begin{figure}
  \centering
  \includegraphics[width = \textwidth, alt={plain-text this plots shows two subplots on top of each other, sharing the same x axis, the thermal energy in the photon bath, $k_B T_\gamma$, from 1e7 eV (where BBN starts) to 1e4 eV (where BBN ends). The upper plot compares the interaction rate from neutron-to-proton and proton-to-neutrino processes to the $M_\phi$ and $3H$, in the decoupled case. The neutron-to-proton rate falls with time but is always larger than $M_\phi$ and $3H$, and saturates to a constant value at 2e5; while the proton-to-neutron falls steeply and becomes effectively zero at 3e5. In the lower plot, the relative decrease that a coupling $g/m_\phi$ produces in the neutron-to-proton interaction rate, for 0.15/eV and 0.5/eV. One can see that the interaction produces a depletion on the interaction rate, which is constant early enough, has a dip around 2e5, and goes to zero at late times. A shaded region shows the depletion that would be produced if the field was constant $\phi = A$, thus showing that the real interaction rate is an average over $\phi$. A gray dotted line is shown for comparison where $m_\nu = 0.12$ MeV, constant: the depletion only happens at late times, and is constant.}]{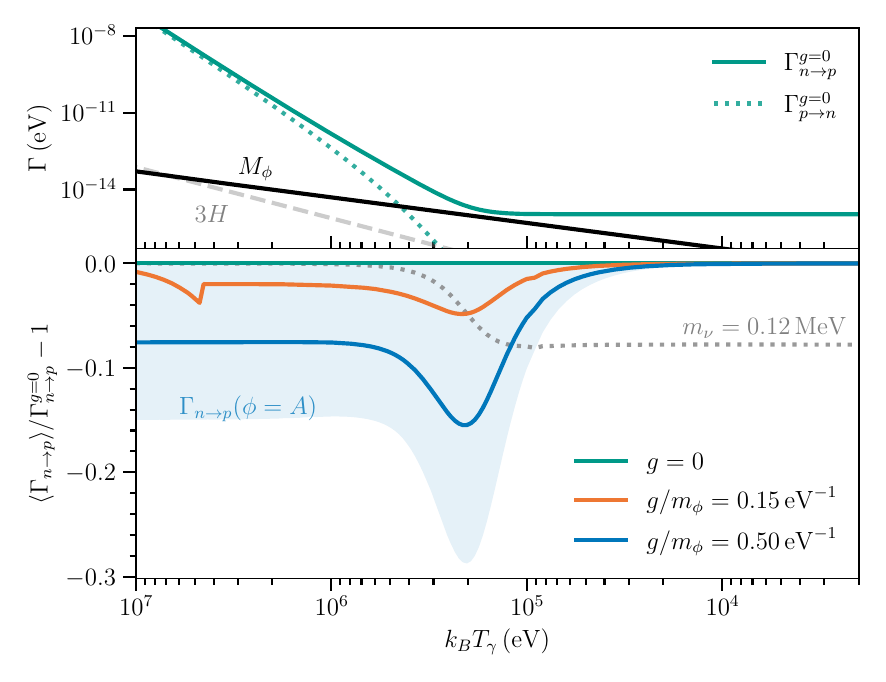}
  \caption{Neutron-to-proton (and viceversa) interaction rates $\Gamma_{n\to p}$ ($\Gamma_{p\to n}$) at the temperatures relevant for BBN, and the effect of $\nu$-ULDM interactions in them. In the top plot, we compare $\Gamma$ to the other relevant timescales of the problem, and check that $3H>\Gamma,M_\phi$; which requires us to average over $\phi$ oscillations, as described in the main text. $M_\phi$ is shown for $g/m_\phi = 0.15\, \mathrm{eV}^{-1}$. In the bottom plot, we see how the coupling reduces the $n\to p$ interaction rate. For comparison, we show the effect of a constant mass $m_\nu = 0.12\, \mathrm{MeV}$ (dotted line). The blue shaded region encloses $\Gamma_{n\to p}(\phi)$ for all possible values of $\phi<A(a)$. Thus, the solid blue line is the average of the blue shaded region. \textit{A larger neutrino mass decreases the interaction rate, which will increase the helium and deuterium abundances.}}
  \label{fig3:bbn-interaction-rate}
\end{figure}

In order to implement neutrino masses in $n\to p$ (and viceversa) interaction rates, we follow the Born approximation, where neutrons and protons are approximated to have infinite mass. In this limit, interaction rates from all processes are given by~\cite{Burns:2023sgx},
\begin{equation}\label{eq3:interaction-rates-Born}
\begin{split}
  \Gamma_{n\to p} &= 
    \tilde G_F^2 \int_0^\infty \dd E_e \, E_e E_\nu^- \sqrt{E_e^2-m_e^2}\sqrt{E_\nu^- -m_\nu^2}\left[f_\nu(E_\nu^-)f_e(-E_e)+f_\nu(-E_\nu^-)f_e(E_e)\right]\, , \\
  \Gamma_{p\to n} &= 
    \tilde G_F^2 \int_0^\infty \dd E_e \, E_e E_\nu^+ \sqrt{E_e^2-m_e^2}\sqrt{E_\nu^+ -m_\nu^2}\left[f_\nu(E_\nu^+)f_e(-E_e)+f_\nu(-E_\nu^+)f_e(E_e)\right]\, .
\end{split}
\end{equation}
Here, $\tilde G_F^2 = G_F V_{\mathrm{ud}}\sqrt{(1+3g_A^2)/(2\pi^3)}$ with $G_F$ the Fermi constant, $V_{\mathrm{ud}}$ the Cabibbo angle and $g_A$ the axial electroweak coupling; and $E_\nu^\pm = E_e \pm \Delta$. 
The distribution functions $f_\nu,\, f_e$ follow Fermi-Dirac distributions as in~\cref{eq3:f0}, controlled by $T_\nu$ and $T_\gamma$, respectively. These implement the amount of allowed fermions for the interaction if in the initial state, or Pauli blocking if in the final state. These rates are shown in \cref{fig3:bbn-interaction-rate}.

Now, $\Gamma = \Gamma(\phi)$ through $m_\nu = m_\nu(\phi)$. Before misalignment, $3H > M_\phi$, the field is frozen and oscillations need not be taken into account. However, after misalignment, the field oscillates and $\Gamma$ needs to be averaged. Different timescales play a role here. First, the oscillation of the field is always much slower than the timescale of the electroweak vertex, $M_\phi^2 \ll G_F^{-1}$, and therefore electroweak interactions happen within a constant value of the field. Then, two regimes can follow,
\begin{itemize}
  \item If $\Gamma > M_\phi$, many interactions happen within an oscillation of the field, and so \cref{eq3:interaction-rates-Born} can reach equilibrium within a constant $m_\nu$. Since $\phi$ oscillates many times within cosmological timescales, we need to average $\Gamma$ over each $m_\nu(\phi)$.
  \item If $\Gamma < M_\phi$, interactions happen only once every many oscillations. At cosmological scale, with $\Gamma > H$, still many interactions happen, each of them effectively at a random value of $m_\nu(\phi)$. Then, the result is still an effective averaged $\Gamma$.
\end{itemize}
In any case, we must average as in \cref{eq3:averaging-oscillation}
\begin{equation}\label{eq3:averaging-rates}
  \langle\Gamma \rangle = t_\phi^{-1}\oint \frac{\Gamma(\phi)}{\dot\phi}\,\dd\phi\, ,
\end{equation}
where $\Gamma(\phi)$ is given by \cref{eq3:interaction-rates-Born} with $m_\nu = m_\nu(\phi)$. Finally, neutrinos do not play a role in the subsequent thermonuclear interactions, and therefore hadronic rate are unmodified by non-zero $m_\nu$. We implement the modified expansion history and expansion rates from \cref{eq3:interaction-rates-Born,eq3:averaging-rates} in the numerical code \texttt{PRyMordial}~\cite{Burns:2023sgx}.
\Cref{fig3:bbn-interaction-rate} shows how the coupling reduces $\langle\Gamma_{n\to p}\rangle$. As a consequence, a reduced $n\to p$ interaction rate will then increase $Y_\mathrm{P}$.

\section{Results and discussion}\label{sec3:pseudo-results}
\begin{figure}[t]
  \centering
  \includegraphics[width = 0.9\textwidth, alt={plain-text two subplots on top of each other show the value of the primordial element abundances as a function of the coupling: helium-4 (above) and deuterium (below); with fixed $m_\phi = 10^{-20}$. In each subplot, a grey band shows the one sigma measured value of the primordial element. For very small couplings, the prediction matches the Standard Model, slightly below the measured value, but in good consistency. For larger couplings, the predicted abundances increase and eventually become much larger than the measured values. A dotted line shows the prediction of the model if only $\Delta N_{eff}$ is considered, and a dashed line only if the increase in $m_\nu$ is considered. For helium, the abundance increase mainly comes from $m_\nu$, while in deuterium both are quite comparable.}]{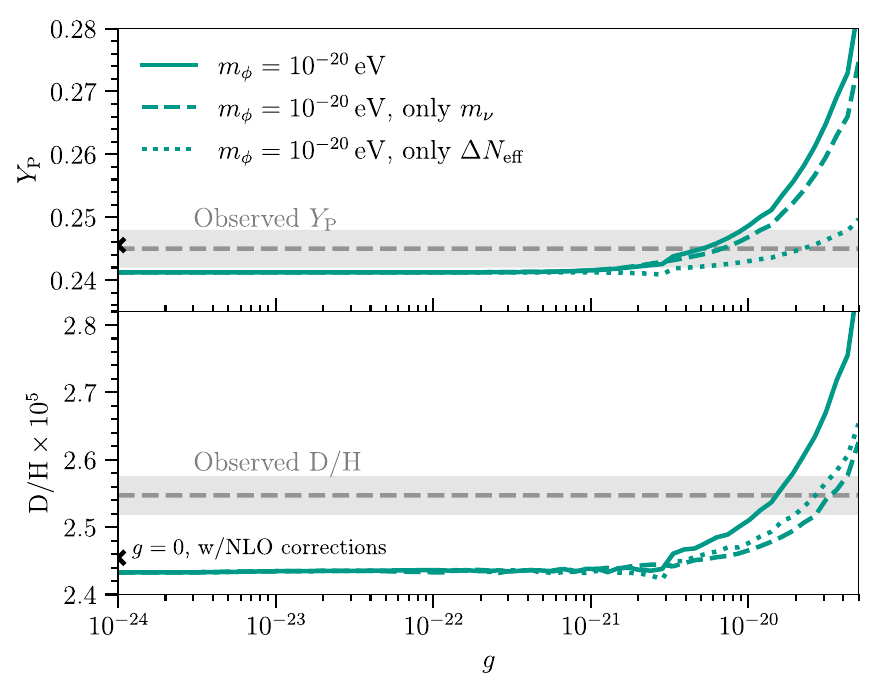}
  \caption{Prediction on primordial abundances of $^4$He and deuterium for different values $g$, at $m_\phi = 10^{-19}\, \mathrm{eV}$. Solid lines show the prediction including all effects of our model, while a dotted line includes only the modification of neutrino masses (i.e., $\Delta N_{\mathrm{eff}}=0$ artificially), and a dashed line only the modification of $N_{\mathrm{eff}}$ (i.e., $m_\nu=0$). A black cross shows the results at $g=0$ beyond the Born approximation, including next-to-leading order contributions, and a grey band shows the $1\sigma$ contours of the observed values~\cite{ParticleDataGroup:2024cfk}.}
  \label{fig3:primordial-abundances}
\end{figure}

The modified version of \texttt{PRyMordial} allows to compute all the light element abundances. 
From all the observed abundances, we discard $^3$He, for which only an upper limit exists, and $^7$Li, which is currently anomalous. 
Thus, restricting the analysis to the abundances given in \cref{eq3:bbn-observed-abundances} makes it more conservative and robust. \Cref{fig3:primordial-abundances} shows how an increased interaction rate increases the expected abundances, far above the observed values.
The results also indicate that, while both $\Delta N_{\mathrm{eff}}$ and $m_\nu$ increase the final abundances, the $Y_{\mathrm{P}}$ result is more sensitive to $m_\nu$ than to $\Delta N_{\mathrm{eff}}$ and is the dominant quantity driving the constraints. Black crosses show the expected $g\to 0$ limit of the full BBN computation, beyond the Born approximation and including next-to-leading order (NLO) corrections. Computation of these corrections in this model is beyond the scope of our work. However, the $g\to 0$ limit shown in \cref{fig3:primordial-abundances} hints that NLO corrections also increase the final abundances, and in this case our bounds can be understood as conservative.

\begin{figure}[t]
  \centering
  \includegraphics[width = 0.9\textwidth, alt ={plain-text a 2d plot shows the parameter space of the model: in the x axis the bare mass of the field $m_\phi$ (between 1e-22 and 1e-17 eV), and in the y axis the coupling $g$ (between 1e-23 and 1e-17). An orange region shows the excluded region from BBN (this work): for $m_\phi$ > 3e-20 eV it increases linearly with m_phi, while for lighter masses it goes roughly as $\sqrt{m_\phi}$. Other bounds are also shown. The bound from naive CMB effects is one order of magnitude weaker, and the bound from flavoured neutrino oscillations is around two orders of magnitude stronger. A gray dashed region for $g > 5e-20$ shows the region where misalignment happens before decoupling, and a gray line much above the bound the region where self-interactions with a scalar field are relevant.}]{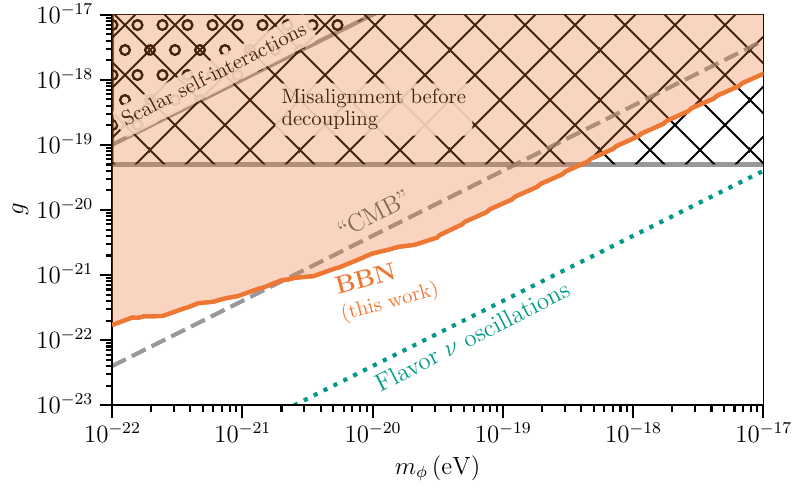}
  \caption{95\% C.L. exclusion regions for the $\nu-\phi$ coupling (as defined in \cref{eq3:lagrangian-pseudo}), for the test statistic defined in~\cref{eq3:pseudo-delta-chi2}. A gray dashed line shows the bound for a scalar coupling to RHNs, from CMB~\cite{Huang:2021kam}. A green dotted line shows a bound for this same model, assuming it has flavor structure and gives rise to time-dependent neutrino oscillations~\cite{Berlin:2016woy}. The region hatched with crosses corresponds to misalignment happening before neutrino decoupling, meaning neutrinos can be non-relativistic at decoupling and modify their distribution function from the here-assumed Fermi-Dirac. The region with circles shows where neutrino self-interactions mediated by a scalar $\phi$ are relevant.}
  \label{fig3:pseudo-money-plot}
\end{figure}

Then, for every point in the $(g,m_\phi)$ parameter space we can define a $\chi^2$ function,
\begin{equation}\label{eq3:pseudo-chi2}
  \chi^2(g,m_\phi) = \left(\frac{Y_{\mathrm{P}}(g,m_\phi)-Y_{\mathrm{P}}^{\mathrm{obs}}}{\sigma_{Y_{\mathrm{P}}}}\right)^2 + \left(\frac{\mathrm{D/H}(g,m_\phi)-(\mathrm{D/H})^{\mathrm{obs}}}{\sigma_{\mathrm{D/H}}}\right)^2\, ,
\end{equation}
where $Y_\mathrm{P}^{\mathrm{obs}}$ ($(\mathrm{D/H})^{\mathrm{obs}}$) is the observed value of the $^4$He ($^2$H) abundance, and $\sigma_{Y_\mathrm{P}}$ ($\sigma_{\mathrm{D/H}}$) its 1$\sigma$ uncertainty, as given in~\cref{eq3:bbn-observed-abundances}. 
We perform a hypothesis test by defining
\begin{equation}\label{eq3:pseudo-delta-chi2}
  \Delta\chi^2 \equiv \chi^2(g,m_\phi) - \chi^2(g=0)\, ,
\end{equation}
where $\chi^2(g=0)$ is the value of \cref{eq3:pseudo-chi2} at $g=0$, i.e., the null hypothesis. 
The 95\% C.L. contours of this function is presented in \cref{fig3:pseudo-money-plot}. At $m_\phi\gtrsim 3\times 10^{-20}\,\mathrm{eV}$, the bound is $g \lesssim 0.13(m_\phi/\mathrm{eV})$. At this region, misalignment happens before BBN, and all observable effects depend only on $g/m_\phi$. However, as explained in \cref{sec3:pseudo-initial-conditions}, bounds for $g \gtrsim 5\times 10^{-20}$ (region hatched with crosses) depend on the physics of decoupling and might be corrected by thermal effects~\cite{Venzor:2020ova,Plestid:2024kyy}. At $m_\phi\lesssim 3\times 10^{-20}\,\mathrm{eV}$, misalignment happens at some point in BBN. Previous to misalignment, the feedback effects disappear and BBN loses constraining power, thus weakening the bound to $g \lesssim 1.8\times 10^{-11}\sqrt{m_\phi/\mathrm{eV}}$. Finally, a region hatched with circles shows where neutrino self-interactions mediated by $\phi$ are relevant for a scalar coupling. Since our bound is below this region, our bound also applies to the scalar coupling.

Due to the Milky Way DM halo, the local DM density is larger than the cosmological mean field value that we have defined up to now. In particular, $\rho^\odot_{\mathrm{DM}} = 0.3\, \mathrm{GeV\, cm}^{-3}$. Taking this into account, it would be specially interesting if this model could predict the right order of magnitude for neutrino masses without requiring the extra parameter $m_0$. Neglecting $m_0$, this means
\begin{equation}
  \left\langle m_\nu^\odot\right\rangle \simeq\frac{g}{m_\phi} \sqrt{\rho^\odot_{\mathrm{DM}}} = 2\times 10^{-4}\left(\frac{g/m_\phi}{0.13\, \mathrm{eV}^{-1}}\right)\, \mathrm{eV}. 
\end{equation}
Setting the target neutrino mass to $m_\nu^\odot\sim \sqrt{|\Delta m_{3\ell}^2|}= 3\times 10^{-3}\, \mathrm{eV}$, in the mass range studied here this would only be reached at couplings saturating the bound for $m_\phi\sim 10^{-22}\, \mathrm{eV}$. 

\section{Conclusions and outlook}\label{sec3:pseudo-conclusions}
The nature of DM and the origin of neutrino masses are two yet unsolved mysteries of particle physics. Models which connect both of these sectors are theoretically motivated and phenomenologically rich. In particular, models of ULDM are gaining interest in the community. In this article, we have explored the phenomenological implications of a coupling between ULDM and the neutrino sector in the early Universe. As a result, we have presented a quantitative bound from a full cosmological evolution of such models.

In this work we have quantitatively solved the equations of motion for the ULDM field, which have a large separation of scales between the frequency of $\phi$ oscillations and cosmological evolution. To such purpose, we follow an adiabatic approximation which provides an invariant quantity with which to compute the amplitude of the field at every time. This method predicts that the scaling of the ULDM field changes when the neutrino interaction potential dominates over its mass potential, from $a^{-3/2}$ to $a^{-1}$. 

Two observable consequences in the early Universe derive from this scaling transition. First, neutrino masses scale linearly with the scale factor, and thus their velocity, related to $T_\nu/m_\nu$, stays constant. This is shown in \cref{fig3:pseudo-neutrino-mass}. When compared to the naive scaling of $a^{-3/2}$, this weakens the effect of time-dependent neutrino masses in the early Universe. Second, the modified scaling also reduces the contribution of the bare-mass potential, $m_\phi^2\phi^2$, to the total energy of the system. As a consequence, the coupled $\nu$-ULDM stays relativistic for a longer time, as in \cref{fig3:pseudo-equation-of-state}. If then we ask for $\phi$ to fulfill the totality of DM, this requires that the radiation energy density in the early Universe is slightly larger, which contributes positively to $N_{\mathrm{eff}}$, as in \cref{fig3:pseudo-neff}. 

Finally, we implement this two phenomenological consequences in Big Bang Nucleosynthesis using the \texttt{PRyMordial} code~\cite{Burns:2023sgx}. We modify the $n\to p$ and $p\to n$ interaction rates to account for a non-zero neutrino mass, following the Born approximation. Since these rates are faster than the oscillation frequencies and the cosmological evolution, we average them for the oscillation of the ULDM field. Using the modified rates and $N_{\mathrm{eff}}$, we compute the primordial element abundances for different $(g,m_\phi)$, as in \cref{fig3:primordial-abundances}. Comparing these to observations, we present our bounds on the parameter space of the model in \cref{fig3:pseudo-money-plot}, which improve CMB bounds by a factor $\sim 4$.

The non-trivial dynamics of the coupled fluid are an unexpected consequence of $\nu$-ULDM which open new research possibilities. In particular, in this work we have focused on BBN consequences since we \textit{a priori} expect it to provide stronger constraints on the model. Still, two arguments motivate using the CMB to further understand and constrain this model. First, the dynamics of $\nu$-ULDM cosmological perturbations could produce unexpected leading-order effects. Second, a joint CMB+BBN analysis would simultaneously infer $\Omega_b$, $g$, and $m_\phi$, potentially modifying constraints by breaking degeneracies between these parameters. However, standard CMB analyses assume that $N_{\rm eff}$ is consistent between BBN and CMB epochs. As \Cref{fig3:pseudo-neff} shows, this is not the case here. This discrepancy motivates a more refined analysis where the relationship between $Y_p$ and $N_{\rm eff}(a_{\rm CMB})$ is adjusted for the model.

The coupling that we have used should be understood as a first approach to $\nu$-ULDM couplings with time-varying masses. Other couplings, which have more theoretically motivated SM extensions, have been proposed in the literature. Even if their phenomenological implications have already been explored~\cite{Plestid:2024kyy,Huang:2021kam}, it would be interesting to extend the quantitative analysis here presented into such models to place consistent bounds into them. 

All in all, cosmology remains an exciting framework within which to look for signatures of BSM particle physics models, specially for particles which interact extremely feebly. That is, the large particle number densities involved compensate for the small couplings. This is specially interesting for interactions which involve the neutrino sector. At the same time, cosmology is a complex science with a lot of subleties that must be under control in order to produce robust statements. In this work we have tried to clarify some of such details and to make claims on $\nu$-ULDM claims consistent and quantitative. 

\section{Acknowledgements}
The authors thank Iván Esteban and Concha González-García for early discussions on neutrino self interactions, and Anne-Katherine Burns for help on the use of \texttt{PRyMordial} and BBN.
This work has been supported by the Spanish MCIN/AEI/10.13039/501100011033 grants PID2022-126224NB-C21 and by the European Union’s Horizon 2020 research and innovation program under the Marie Skłodowska-Curie grants HORIZON-MSCA-2021-SE-01/101086085-ASYMMETRY and H2020-MSCA-ITN-2019/860881-HIDDeN, and by the ``Unit of Excellence Maria de Maeztu 2020-2023'' award to the ICC-UB CEX2019-000918-M (TB and JS). TB is supported by the Spanish grant PRE2020-091896. JLS is supported by Grants PGC2022-126078NB-C21 funded by MCIN/AEI/10.13039/501100011033 and ``ERDF A way of making Europe'' and Grant DGA-FSE grant 2020-E21-17R Aragon Government and the European Union-NextGenerationEU Recovery and Resilience Program on ``Astrofísica y Física de Altas Energías'' CEFCA-CAPA-ITAINNOVA.

\bibliographystyle{JHEP}
\bibliography{biblio.bib}

\clearpage
\appendix

\section{Details on the formalism}\label{app:formalism}
In this appendix, we formalize definitions for variables and functions defined in the main text.

\subsection{Solutions to the equations of motion}
Firstly, \cref{eq3:expectation-value} introduces the classical expectation value for a certain quantum operator $\hat{\mathcal{O}}$, 
\begin{equation}
  \langle\hat{\mathcal{O}}\rangle = 
  \sum_s \int \dd P_1\dd P_2 \dd P_3 \frac{1}{\sqrt{-\det g}}
  \frac{1}{2P^0} f(P,x,s) \langle\phi,P^s| \hat{\mathcal{O}}|\phi,P^s\rangle\, .
\end{equation}
Here, $|\phi,P^s\rangle = |\phi\rangle \otimes|P^s\rangle$ is the product of a coherent state with field $\phi$ and a one-particle fermion state, 
given by~\cite{Sudarshan:1963ts,Glauber:1963tx,Glauber:1963fi,Peskin:1995ev}
\begin{equation}
  \begin{split}
    |\phi\rangle &  \equiv 
    \exp\left\{-\frac{1}{2} \int \frac{\dd^3 \vec{k}}{(2 \pi)^3} \frac{|\phi(K)|^2}{\left(2 K^0\right)^5}\right\} \,
    \exp\left\{\int \frac{\dd^3 \vec{k}}{(2 \pi)^3} \frac{\phi(K)}{\left(2 K^0\right)^{5 / 2}} a_K^{\phi^{\dagger}}\right\}\,|0\rangle, \\ 
    |P^s\rangle & = \sqrt{2P^0}{a_P^s}^\dagger|0\rangle\, . 
  \end{split}
\end{equation}
Here, $|0\rangle$ is the vacuum state, $a_P^s$ ($a^\phi_K$) is the annihilation operator of the field $\psi$ ($\phi$), $K^\mu = (E_K, \vec{k})$ such that $K^2 = -m_\phi^2$, and $\phi(K)$ is the Fourier transform of the classical scalar field $\phi(x)$,
\begin{equation}
  \phi(x) \equiv \int \frac{\dd^3 \vec{k}}{(2 \pi)^3} \frac{1}{(\sqrt{2} K^0)^3}\left[\phi(K) e^{-i K x}+\phi(K)^* e^{i K x}\right]\, .
\end{equation}
Then, $P_\mu$ is the conjugate momentum to $x^\mu$, and is related to the 4-momentum of the particle $P^\nu$ by $P_\mu = g_{\mu\nu}P^\nu$, and to the physical momentum $p^i = p_i$ as $P_i = ap_i$. The 4-momentum of the particle is described by the geodesic equation
\begin{equation}\label{eq3:geodesic-eq}
  P^0\frac{\dd P^\mu}{\dd \eta} + \Gamma^\mu_{\ \nu\rho}P^\nu P^\rho = 0\, .
\end{equation}
In a homogeneous Universe, this returns that the physical linear momentum, $p_i = P_i/a$, decreases as $a^{-1}$ as the Universe expands. In this work, we have worked with the comoving linear momentum, $q_i = a p_i$, which we describe in terms of its magnitude and direction, $q_i = qn_i$.

The general solution to the Dirac equation, \cref{eq3:eom-fermion-pseudo}, is
\begin{equation}\label{eq3:psi-plane-wave-decomposition}
  \begin{split}
    \psi &= \int \dpppE \sum_{s=1,2} \left(e^{-iPx}a^s(\vec{p})u^s(\vec{p})+e^{iPx}{b^s}^\dagger(\vec{p})v^s(\vec{p})\right)\, ,
  \end{split}
\end{equation}
with 
\begin{equation}
  u^s(\vec{p}) = e^{i\alpha \gamma^5}u^s_0(\vec{p})\,,\qquad
  v^s(\vec{p}) = e^{i\alpha \gamma^5}v^s_0(\vec{p})\, ,
\end{equation}
the plane wave solutions, $P^\mu = (E_p,\vec{p})$ and $a^s(P)$ and $b^s(P)$ are the annihilation operators to $u^s(\vec{p})$ and $v^s(\vec{p})$. 
For instance, one can use the Pauli-Dirac representation of the gamma matrices,
\begin{equation}
  \gamma^0 = \begin{pmatrix} \mathbb{I}_2 & 0  \\ 0 & -\mathbb{I}_2 \end{pmatrix}\, ,\quad
  \gamma^i = \begin{pmatrix} 0 & \sigma^i \\ -\sigma^i & 0 \end{pmatrix}\, , \quad
  \gamma^5 = \begin{pmatrix} 0 & \mathbb{I}_2 \\ \mathbb{I}_2 & 0 \end{pmatrix}\, ,
\end{equation}
with $\mathbb{I}_2$ the 2x2 identity matrix, and $\sigma^i$ the Pauli matrices. Then, the positive and negative frequency solutions to \cref{eq3:eom-fermion-diagonalised}, with defined mass $m_\nu$, are~\cite{Halzen:1984mc} 
\begin{equation}\label{eq3:dirac-solutions}
  u^s_0(\vec{p}) = \begin{pmatrix} 
    \sqrt{E_p+m_\nu}\, \xi^s \\ \dfrac{\vec\sigma\cdot\vec{p}}{\sqrt{E_p+m_\nu}}\, \xi^s 
  \end{pmatrix}\, , \qquad 
  v^s_0(\vec{p}) = \begin{pmatrix} 
    \dfrac{\vec\sigma\cdot\vec{p}}{\sqrt{E_p+m_\nu}}\, \chi^{s} \\ \sqrt{E_p+m_\nu}\, \chi^{s} \end{pmatrix}\, ,
\end{equation}
respectively. Here $E_p = \sqrt{\vec{p}^2 + m_0^2 + g^2\phi^2}$, and $\xi^s,\,\chi^s$ are spinors corresponding to the eigenvectors of a Stern-Gerlach experiment in an arbitrary $\hat{n}$ direction, with $s=1,\,2$ that differentiates between the two helicity eigenstates. These must fulfill ${\xi^s}^\dagger\xi^s = {\chi^s}^\dagger\chi^s =1$.
With all this information, one can check that the plane wave solutions fulfill
\begin{equation}
  \begin{split}
    \bar{u}^s(\vec{p})\gamma^0 u^r(\vec{p}) = 
    \bar{v}^s(\vec{p})\gamma^0 v^r(\vec{p}) &= 2 E\delta^{sr} \, , \\
    \bar{u}^s(\vec{p})\gamma^5 u^r(\vec{p}) = 
    -\bar{v}^s(\vec{p})\gamma^5 v^r(\vec{p}) &= 2ig\phi\, \delta^{sr}\, .
  \end{split}
  \end{equation}

\subsection{Evolution of the distribution function}
In general, the neutrino phase space distribution $f(x^i, q, n_j, \eta)$ evolves according to the Boltzmann equation
\begin{equation}\label{eq3:Boltzmann-equation}
  \frac{D f}{\dd \eta} = 
    \frac{\partial f}{\partial \eta} + 
    \frac{\dd x^i}{\dd \eta} \frac{\partial f}{\partial x^i} + 
    \frac{\dd q}{\dd \eta}\frac{\partial f}{\partial q} + 
    \frac{\dd n_i}{\dd \eta}\frac{\partial f}{\partial n_i} = 
    \frac{\partial f}{\partial \eta}\biggr|_C\, .
\end{equation}
Here, the right-hand side describes the changes on the distribution function due to collisions. For a decoupled ($g=0$) species with constant mass in a homogeneous Universe, $\dd q/\dd\eta = 0$, $\dd n_i/\dd\eta = 0$ and $\partial f/\partial x^i = 0$, which leads to $\partial f/\partial\eta = 0$. This means that any distribution function which depends on $q=pa$ is a solution to the homogeneous collisionless Boltzmann equation. That is, $f(x^i,P_j,\eta) = f_0(q)$, which will maintain its shape and only have the physical momenta redshifted by $p = q/a$. 

However, our model includes mass-varying neutrinos with $m_\nu(\eta) = \sqrt{m_0^2+g^2\phi(\eta)^2}$, which might make us wonder if $\dd q/\dd\eta = 0$ as we had assumed in the standard case. To first order, we would then have
\begin{equation}
  \frac{\partial f_0}{\partial\eta} + \frac{\dd q}{\dd\eta}\frac{\partial f_0}{\partial q} = 0\, ,
\end{equation}
which could make $f_0(q)$ vary with time. However, for mass-varying neutrinos, the geodesic equation gets modified to~\cite{Brookfield:2005bz}
\begin{equation}
  P^0\, \frac{\dd P^\mu}{\dd\eta}+ \Gamma^\mu_{\ \nu\rho}P^\nu P^\rho = 
  -m_\nu^2\, \frac{\dd \log m_\nu}{\dd\phi}\frac{\partial\phi}{\partial x_\mu}\, .
\end{equation}
After some algebra, in the 0th component of the geodesic equation the contribution from $\phi$ vanishes and we still get
\begin{equation}
  \frac{\dd q}{\dd \eta} = 0\, .
\end{equation}
Therefore, $\partial f_0/\partial\eta = 0$ and the shape of the momentum distribution function $f_0(q)$ does not vary with time, even for the mass-varying neutrinos from this model. 

\subsection{Derivation of the adiabatic invariant}\label{app:adiabatic}
\Cref{sec3:pseudo-adiabatic} introduced the adiabatic approximation and obtained the adiabatic invariant $\mathcal{I}(a)$. In this Section we describe the computations to arrive to this result. 

We are interested in obtaining the evolution equation for $\rho$. To do so, we begin by multiplying~\cref{eq3:eom-pseudoscalar-tau} by $\dot\phi$, and rewriting it into
\begin{equation}\label{eq3:adiabatic-drhodt}
  \frac{\dd \rho(a,t)}{\dd t} = H\left[\frac{\partial \rho_\nu}{\partial\log a}-3\dot\phi^2\right]\, ,
\end{equation}
Within cosmological timescales, the energy density varies as 
\begin{equation}
  H\frac{\dd \rho(a,t)}{\dd\log a} = \frac{\dd\rho(a,t)}{\dd t}\, .
\end{equation}
In the adiabatic approximation, we replace $\dd\rho/\dd t$ by its averaged value over an oscillation of the field
\begin{equation}
  H\frac{\dd \rho}{\dd\log a} = \left\langle\frac{\dd\rho}{\dd t}\right\rangle\, .
\end{equation}
Now, plugging~\cref{eq3:adiabatic-drhodt} here,
\begin{equation}\label{eq3:adiabatic-drhodloga}
  \frac{\dd\rho}{\dd \log a} = 
  \left\langle \frac{\partial \rho_\nu}{\partial\log a}\right\rangle - 3
  \langle \dot\phi^2\rangle\ .
\end{equation}

Within an oscillation we can treat the total energy as a constant of motion, $\rho(a,t) \simeq \langle\rho(a,t)\rangle \equiv \rho(a)$. 
This allows to use~\cref{eq3:adiabatic-rho} to find a closed expression for $\dot\phi$,
\begin{equation}
  \dot\phi = \sqrt{2}\sqrt{\rho(a) - m_\phi^2\phi^2/2 - \rho_\nu(\phi) }
\end{equation}
Then,
\begin{equation}
  t_\phi\langle\dot\phi^2 \rangle = \oint\dd \phi\, \dot\phi = 
  \sqrt{2}\oint\dd \phi\, \sqrt{\rho(a) - \frac{1}{2}m_\phi^2\phi^2 - \rho_\nu(\phi)}\, .
\end{equation}
This is an important quantity, since we notice that
\begin{equation}
  \begin{split}
    t_\phi\frac{\dd\rho}{\dd\log a} - t_\phi \left\langle \frac{\partial \rho_\nu}{\partial\log a}\right\rangle &= 
    \frac{1}{\sqrt{2}}\oint\,  \dd\phi \frac{\partial(\rho(a)-\rho_\nu(a,\phi))/\partial\log a}{\sqrt{\rho(a) - \frac{1}{2}m_\phi^2\phi^2 - \rho_\nu(\phi)}} 
    = \\ &=
    \sqrt{2}\, \frac{\dd}{\dd\log a}\oint\dd \phi\, \sqrt{\rho(a) - \frac{1}{2}m_\phi^2\phi^2 - \rho_\nu(\phi)} 
    = \\ &=
    \frac{\dd }{\dd\log a} \left(t_\phi\langle\dot\phi^2\rangle\right)\, .
  \end{split}
\end{equation}
Plugging this result into~\cref{eq3:adiabatic-drhodloga}, we get that
\begin{equation}
  \frac{\dd }{\dd\log a} \left(t_\phi\langle\dot\phi^2\rangle\right) = -3t_\phi\langle\dot\phi^2\rangle\, .
\end{equation}
So, 
\begin{equation}
  \mathcal{I}(a) \equiv \frac{1}{\sqrt{2}} a^3t_\phi\langle \dot\phi^2\rangle = 
  a^3\oint\dd \phi\, \sqrt{\rho(a) - \frac{1}{2}m_\phi^2\phi^2 - \rho_\nu(\phi)}
\end{equation}
is an invariant quantity for the whole cosmological evolution, since $\dd\mathcal{I}/\dd a = 0$. This adiabatic invariant allows to compute the total energy density $\rho(a)$ of the system as a function of the scale factor, given some initial conditions.

\end{document}